\documentclass[%
 reprint,
 amsmath,amssymb,
 aps,
]{revtex4-1}
\usepackage{amsthm}
\usepackage{graphicx}% Include figure files
\usepackage{dcolumn}% Align table columns on decimal point
\usepackage{bm}% bold math
\usepackage{hyperref}% add hypertext capabilities
\usepackage{dsfont}
\usepackage[caption=false]{subfig}
\usepackage{float}

\usepackage[caption=false]{subfig}
\usepackage{algorithm}
\usepackage{algpseudocode}

\newcommand{\GG}{\mathcal{G}}

\newcommand{\E}{\mathcal{E}}

\newcommand{\ones}{\mathbf{1}}

\newcommand{\Mod}{\operatorname{Mod}}

\begin{document}

\preprint{APS/123-QED}

\title{Network clustering and community detection using\\modulus of families of loops}% Force line breaks with \\
\thanks{This material is based upon work supported by the National Science Foundation under Grants No. DMS-1515810}
\email{Corresponding author: heman@ksu.edu} 
\author{Heman Shakeri$^{1}$}
\author{Pietro Poggi-Corradini$^2$}
\author{Nathan Albin$^2$}%
\author{Caterina Scoglio$^1$}
\affiliation{$^1$Electrical and Computer Engineering Department, Kansas State University, Manhattan, Kansas, USA}\affiliation{$^2$Mathematics Department, Kansas State University, Manhattan, Kansas, USA}
\date{\today}% It is always \today, today,

\begin{abstract}
We study the structure of loops in networks using the notion of modulus of loop families. We introduce a new measure of network clustering by quantifying the richness of families of (simple) loops. Modulus tries to minimize the expected overlap among loops by spreading the expected link-usage optimally. We propose weighting networks using these expected link-usages to improve classical community detection algorithms. We show that the proposed method enhances the performance of certain algorithms, such as spectral partitioning and modularity maximization heuristics, on standard benchmarks.
\end{abstract}

\maketitle
%---------------========================-------------------===========
\section{Introduction}
%Structure of loops in networks has 
 
Real networks contain closely connected subnetworks with local structural patterns characterized by their richness of loop \cite{milo2002network}. Loops offer more pathways within them compared to treelike topologies; thus rich loop structures improve network robustness \cite{mugisha2016identifying} and impact propagating and transporting processes in networks \cite{petermann2004role}. Previous approaches on analysis of loop structures focus on loops with lengths of order $3$--$5$ separately \cite{newman2003structure, lind2005cycles} and few such as \cite{bianconi2003number, kim2005cyclic} emphasize the role of higher order loops to characterize their overall structures. 
 We consider assessing loop structures in the network, with any order and altogether and apply our tool for analyzing network transitivity known as clustering coefficient and providing more information for community detection algorithms.

%Real and randomized networks differ in the existence of statistically significant subnetworks, i.e. motifs \cite{milo2002network}; highlighting the role of closely connected structural patterns such as loops. Compared to treelike topologies, loops offer more pathways within them; thus, rich loop structures impact the propagating and transporting processes in networks \cite{petermann2004role}.

Our goal is to study loop structures in the network using the concept of modulus of loop families developed in \cite{Modulus},  \cite{Albin2014}, and \cite{albin2016minimal}. Modulus is a way of measuring the richness of certain families of objects on a network, such as loops, walks, trees, etc,  and  is a discrete analog of the classical theory of modulus of curve families in complex analysis \cite{ahlfors1973}. Although modulus on networks is not a new concept (see \cite{duffin:1962jmaa} and \cite{schramm:1993israel}), it is not as well developed as in the continuum setting.
In \cite{Modulus}, the authors showed that modulus is a standard convex optimization problem.
Continuity and smoothness 
properties of modulus on networks were considered in \cite{Albin2014}.  A probabilistic interpretation provided in \cite{albin2016minimal}.

Modulus is a versatile tool to analyze networks. 
Different types of families of walks can be used to learn about different aspects of the network. In \cite{shakeri2016generalized}, we introduced centrality measures based on various families of walks that can be computed on directed or undirected, weighted or unweighted, and even disconnected networks. 
These measures do not necessarily have to consider the whole network. We applied them to detect influential sections of the network, ranking the nodes, and we explored applications to improve vaccination strategies for reducing the risk of epidemics. The applications to epidemic spreading were further studied in \cite{goering2015numerical}, where the authors used modulus 
to analyze the concept of Epidemic Hitting Time.

Our main contributions in this paper are introducing a generic approach to analyze loops structures in the network that consider local loop topologies with an eye on the entire network. We quantify richness of loops and introduce a clustering measure based on that. Moreover, we find the probablity of usage of each link in important loops and use it as a  measure of affinity between nodes to enhance network partitioning.

This paper is organized as follows. First, we introduce our notation and the necessary background on modulus of families of loops. Then, we define our proposed methods to measure clustering in the network. Next, we show how to preprocess a network in order to improve partitioning techniques such as Fiedler vector bisection and the modularity maximization heuristics. Finally, we discuss other potential applications.

\section{Notations and Definitions}\label{sec:notation}
Let $\GG=(V,E)$ be a network with nodes $V$ and links $E$.
A \emph{walk} is a string of nodes $\gamma = v_0 v_1 \cdots v_n$ on $\GG$ with the property that consecutive nodes $v_i$ and $v_{i+1}$ are linked in the network. 
%We use $\Gamma$ to represent an arbitrary nonempty family of walks on $\GG$.
A walk $\gamma=v_1 v_2 v_3 \ldots v_r$, is a {\it simple loop} if the nodes $v_i$ are all distinct, except that $v_r=v_1$. We call $\mathcal{L}$ the family of all loops in $\GG$. Other possible loop families are loop families rooted at a given node $v$ or link $e$; we write $\mathcal{L}^v$ or $\mathcal{L}^e$ in that case.

Given a density $\rho:E\rightarrow \left[0,\infty\right)$, interpreted as a penalty or cost the walker must pay for traversing link $e$,
we define the {\it $\rho$-length} of a loop $\gamma$ as
\begin{equation}
\ell_\rho\left(\gamma\right):=\sum_{e\in \gamma}\rho\left(e\right).
\end{equation}
When $\rho_0\left(e\right)\equiv 1$, then $\ell_{\rho_0}(\gamma)$ represents the hop-length of $\gamma$. Likewise, given a family of loops $\mathcal{L}$ we set $\ell_\rho\left(\mathcal{L}\right):=\min_{\gamma\in\mathcal{L}}\ell_\rho\left(\gamma\right)$.
We introduce a $|\mathcal{L}|\times|E|$ matrix $\mathcal{N}$ such that each row corresponds to a loop $\gamma\in\mathcal{L}$ and is the indicator function $\mathds{1}_{e\in\gamma}$.

Let $w:E\rightarrow \left(0,\infty\right)$ be a positive weight function. Then, for $1< p<\infty$, $\Mod_{p,w}\left(\mathcal{L}\right)$ is defined as 
\begin{equation}\label{eq_modulus0}
\Mod_{p,w}\left(\mathcal{L}\right) = \min_{\lbrace \rho|\ell_\rho(\mathcal{L})>0\rbrace} \frac{\E_{p,w}}{\ell_\rho(\mathcal{L})^p}
\end{equation} 
where $\E_{p,w}(\rho) =\sum_{e\in E} w\left(e\right)\rho\left(e\right)^p$
is the energy of the density $\rho$.
% and $\rho^*$ is the unique minimizer \cite[Lemma 2.1]{Albin2014}. 
 In this paper, we work with an equivalent form of \eqref{eq_modulus0} defined as in \cite{Modulus}:

\begin{equation}\label{eq:modulus}
\Mod_{p,w}\left(\mathcal{L}\right) = \min_{\lbrace \rho| \mathcal{N}\rho \geq 1\rbrace}\E_{p,w}\left(\rho\right)=\E_{p,w}\left(\rho^\ast\right),
\end{equation}

We call a density $\rho$ with $\mathcal{N}\rho \geq 1$ admissible $\rho$ for a family of loops $\mathcal{L}$.

  For example, if $\GG$ is a tree, $\Mod_p\left(\mathcal{L}\right)=0$ by Property (d) below; if $\GG$ is an unweighted complete graph, then $\Mod_p\left(\mathcal{L}\right)=\frac{1}{3^p}{n\choose 2}$.

For a finite network $\GG$, the following properties hold, see \cite{Modulus,shakeri2016generalized}:
\begin{itemize}
\item[(a)] {\rm \textbf{$\mathbf{p}$-Monotonicity:}}  The extremal densities satisfy $0\leq \rho^*\left(e\right) \leq 1$ for all $e\in E$. Thus, for $1\le p\leq q$, we have $\Mod_q\left(\mathcal{L}\right)\leq \Mod_p\left(\mathcal{L}\right)$.

\item[(b)] {\rm \textbf{$\mathbf{\mathcal{L}}$-Monotonicity:}} If $\mathcal{L}'\subset \mathcal{L}$, then $\Mod_p\left(\mathcal{L}'\right)\leq \Mod_p\left(\mathcal{L}\right)$.

\item[(c)] {\rm \textbf{$\mathbf{w}$-Monotonicity:}} If $w$ and $w'$ are positive link weights with $w\le w'$ then $\Mod_{p,w}(\mathcal{L})\le\Mod_{p,w'}(\mathcal{L})$.

\item[(d)] {\rm \textbf{Empty Family:}} If $\mathcal{L} = \emptyset$, then $\Mod_p\left(\mathcal{L}\right)=0$.

\item[(e)] {\rm \textbf{Countable Subadditivity:}} For any sequence $\{\mathcal{L}_i\}_{i=1}^\infty$ of families of loops, $$\Mod_p\left(\cup_{i=1}^\infty\mathcal{L}_i\right)\leq \sum_{i=1}^\infty \Mod_p\left(\mathcal{L}_i\right).$$

\end{itemize}

The properties above allow quantification of the richness of various family of loops, i.e., a family with many short loops has a larger modulus than a family with fewer and longer loops. In particular, $\mathcal{L}$-monotonocity and subadditivity often define a notion of capacity on the set of loops in a network.
For the rest of this paper, we consider $p=2$ due to its physical and probabilistic interpretations as well as computational advantages, for instance, in this case \eqref{eq:modulus} is a quadratic program.

\subsection{Interpreting loop modulus as a measure of the richness of a family of loops}
In order to measure the richness of a family of loops, we want to balance the number of different loops with relatively little overlap vs. how many short loops there are in the family. 

\begin{figure}[h]
\subfloat[]{%
  \includegraphics[clip,width=.35\columnwidth]{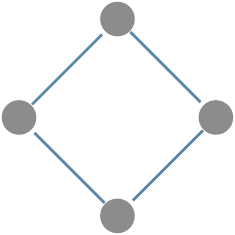}%
}~~~
\subfloat[]{%
  \includegraphics[clip,width=.35\columnwidth]{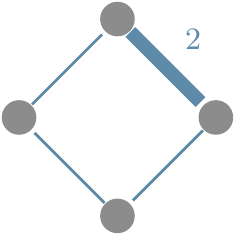}%
}\\
\subfloat[]{%
  \includegraphics[clip,width=.35\columnwidth]{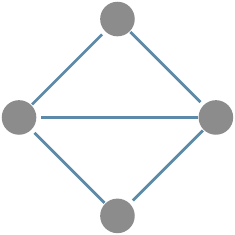}%
}~~~
\subfloat[]{%
  \includegraphics[clip,width=.35\columnwidth]{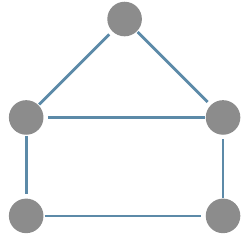}%
}
\caption{Loop Modulus for some networks  demonstrating how modulus can quantify the richness of loops, 
a) $\Mod_2\left(\mathcal{L}\right) = 0.25$
b) Weight of a link is doubled, modulus increase by $w$-monotonicity:  $\Mod_2\left(\mathcal{L}\right) = 0.285$
c) Increasing number of short loops the modulus increases by $\mathcal{L}$-monotonicity: $\Mod_2\left(\mathcal{L}\right) = 0.5$.
d) Loops are longer than (c) and modulus decreases: $\Mod_2\left(\mathcal{L}\right) = 0.455$.}\label{fig:Mods}
\end{figure}

We demonstrate this in Figure \ref{fig:Mods}.
For the square in Figure \ref{fig:Mods}(a), the family $\mathcal{L}$ consists of a single loop, hence  $\Mod_2\left(\mathcal{L}\right) = 0.25$.
 In Figure \ref{fig:Mods}(b), the weight of one link is doubled and modulus increases to $\Mod_2\left(\mathcal{L}\right) = 0.285$, as it must,  by $w$-monotonicity (Property~(c)). The network in Figure \ref{fig:Mods}(c) has more loops than the one in Figure \ref{fig:Mods}(a) and modulus increases to $\Mod_2\left(\mathcal{L}\right) = 0.5$, demonstrating $\mathcal{L}$-monotonicity (Property~(b)). 
 Comparing Figure \ref{fig:Mods}(c) to Figure \ref{fig:Mods}(d), we see that they have the same number of loops, but in (d) they are longer and thus the modulus decreases to $\Mod_2\left(\mathcal{L}\right) = 0.455$.

\subsection{Probability interpretation of loop modulus}\label{sec:probInterp}
For $p=2$ the modulus problem in  \eqref{eq:modulus} is
\begin{equation}\label{eq_primal}
\min_{\substack{\lbrace \rho|\mathcal{N}\rho\geq \ones\rbrace}} \rho^T\rho.
\end{equation}
We consider the Lagrangian for \eqref{eq_primal}:
\begin{equation}
L(\rho, \lambda) = \rho^T\rho -\lambda^T\left( \mathcal{N}^T \rho -\ones\right),
\end{equation}
where $\lambda \in \mathbb{R}^\mathcal{L}_{\ge 0}$ is the Lagrange multipliers. 
It is easy to show that $\rho = \ones$ is an interior point for the feasible region of \eqref{eq_primal}, thus strong duality holds (Slater's condition \cite{boyd2004convex}).
% and:
%\begin{equation}\label{eq_StrongDuality}
%\Mod_2(\mathcal{L}) = \sup_{\lambda\ge 0}\inf_\rho L(\rho, \lambda) 
%\end{equation}
Minimizing $L$ in $\rho$  gives   
\begin{equation}\label{eq_rho_opt}
\rho^*(e) = \frac{1}{2} \sum_{\gamma \in \mathcal{L}} \lambda^*(\gamma) \mathds{1}_{e\in \gamma},
\end{equation}
and the  dual problem: 
\begin{equation}\label{eq_Dual}
\max_{\substack{\lambda\geq 0}} \left(\lambda^T \ones  - \frac{1}{4}\lambda^T C \lambda\right).
\end{equation}
where $C$ is the {\it overlap matrix} for $\mathcal{L}$. Namely, 
\[
C(\gamma_i,\gamma_j) = 
\sum_{e\in E} \mathcal{N}(\gamma_i,e)\mathcal{N}(\gamma_j,e)=|\gamma_i\cap\gamma_j|
\] 
measures the overlap of two loops.

We define a probability mass function  $\mu \in \mathcal{P}(\mathcal{L}):=\lbrace \mu \in \mathbb{R}^\mathcal{L}_{\ge 0} :\mu \ones = 1\rbrace$ that defines a random loop $\underline{\gamma}\in \mathcal{L}$ with
\begin{equation}\label{eq_ProbMu}
\mu(\gamma) = \Pr(\underline{\gamma} = \gamma).
\end{equation}

Writing $\lambda=\nu\mu$ for a nonnegative scalar $\nu$ and a pmf $\mu$ \eqref{eq_Dual} becomes:

\begin{equation}\label{eq_Dual2}
\max_{\substack{\nu\geq 0}} \left(\nu  - \frac{\nu^2}{4}\min_{\mu\in \mathcal{P}(\mathcal{L})}\mu^T C \mu\right).
\end{equation}
The maximum in \eqref{eq_Dual2} occurs when
\begin{equation}\label{eq_nustar}
\nu^* = 2\left(\min_{\mu\in \mathcal{P}(\mathcal{L})}\mu^T C \mu\right)^{-1}
\end{equation}
%The strong duality in  \eqref{eq_StrongDuality} becomes
%\begin{equation}\label{eq_StrongDuality2}
%\Mod_2(\mathcal{L}) = \max_{\substack{\nu\geq 0}} \left(\nu  - \frac{\nu^2}{4}\min_{\mu\in \mathcal{P}(\mathcal{L})}\mu^T C \mu\right).
%\end{equation}
Substituting \eqref{eq_nustar} in \eqref{eq_Dual2}, we get that $\nu^*  = 2\Mod_2(\mathcal{L})$ and 
\[
\Mod_2(\mathcal{L})^{-1} = \min _{\mu\in \mathcal{P}(\mathcal{L})}\mu^T C \mu =\mathbb{E}_{\mu^*}\left|\underline{\gamma_i}\cap\underline{\gamma_j}\right|,
\]
for an optimal $\mu^*$, where $\mathbb{E}_{\mu^*}\left|\underline{\gamma_i}\cap\underline{\gamma_j}\right|$ is the minimum expected overlap of two independent, identically distributed random loops  with pmf $\mu^*\in\mathcal{P}(\mathcal{L})$.

Moreover by \eqref{eq_rho_opt}, the exremal density satisfies
\[
\rho^*(e) =  \Mod_2(\mathcal{L})\mathbb{E}_{\mu^*} \left[\mathcal{N}(\underline{\gamma}, e) \right]
\]
where $\mathbb{E}_{\mu^*} \left[\mathcal{N}(\underline{\gamma}, e )\right]= \sum_{\gamma\in \mathcal{L}} \mathcal{N}(\gamma , e)\mu^*(\gamma) $ is the expected usage of link $e$ in loop $\underline{\gamma}$. Therefore, the optimal measures $\mu^*$ are related to the optimal density $\rho^*$ as follows:
\begin{equation}\label{eq:probInterp}
\frac{\rho^*(e)}{\Mod_2(\mathcal{L})} = \mathbb{P}_{\mu^*}\left(e\in \underline{\gamma}\right)
\end{equation}
We call $\mathbb{P}_{\mu^*}\left(e\in \underline{\gamma}\right)$ the {\it expected usage} of link $e$.

Moreover, one can always find an optimal measure $\mu^*$ that is supported on a minimal set of loops of cardinality bounded above by $|E|$, see \cite[Theorem 3.5]{albin2016minimal}. We think of these loops as ``important loops'' that play a role in the optimization problems as active constraints.

\subsection{Approximating the modulus}
The numerical results in the examples that follow are produced by a Python implementation of the simple algorithm described in \cite{Modulus}.  This algorithm exploits the $\mathcal{L}$-monotonicity (Property~(b)) of the modulus by building a subset $\mathcal{L}'\subseteq\mathcal{L}$ so that $\Mod_2(\mathcal{L}')\approx\Mod_2(\mathcal{L})$ to a desired accuracy \cite[Theorem~9.1]{Modulus}.  In short, the algorithm begins with $\mathcal{L}'=\emptyset$, for which the choice $\rho\equiv 0$ is optimal and insert a loop with the shortest hop-length then repeatedly adds violated constraints to $\mathcal{L}'$ and determines the optimal $\rho$ each time.  The algorithm terminates when all constraints are satisfied to a given tolerance (Algorithm 1).

%------------------------------------------Alg------------
\begin{figure}
\begin{minipage}{\linewidth}
\begin{algorithm}[H]
  \caption{Approximating densities for $\Mod_2(\mathcal{L})$ with tolerance $0<\epsilon_\text{tol}<1$ \cite{Modulus} }
  \label{alg:DS}
   \begin{algorithmic}[1]
   \State  $\rho\leftarrow 0$; $\rho_0\leftarrow \ones$
   \State $\mathcal{L}'\leftarrow \emptyset$
%   \State $\gamma\leftarrow \textsl{find a violating constraint with }(\ell(\rho)\rho_0)$  
 \State $\gamma\leftarrow \textsl{ShortestLoop}(\rho_0)$   
   \While{$\exists \gamma$ such that $\ell_{\rho}(\gamma)\leq 1-\epsilon_\text{tol}$}
		\State $\mathcal{L}'\leftarrow \mathcal{L}'\cup \lbrace \gamma \rbrace$
		\State $\rho\leftarrow\text{argmin} \lbrace \E_2(\rho):\mathcal{N}\rho \geq \ones\rbrace$
	    		
   \EndWhile
   \end{algorithmic}
\end{algorithm}
\end{minipage}
\end{figure}
%-------------------------------------End of Alg------------

The two key ingredients for implementing this algorithm are a solver for the convex optimization problem \eqref{eq:modulus} and a method for finding violated loops, i.e., with $\rho$-length less than one.  In our implementation, the optimization problem is solved using an active set quadratic program \cite{goldfarb1983numerically} and the violated constraint search is performed using a modified version of the breadth-first search from each node that has a cut-off $1-$tol and reports the first backward link that forms a loop less than the cut-off.

%In practice we only need to find a loop $\gamma$ that has length $\ell(\gamma) \le 1-$tol and is violating the admissiblity constraint thus we add it to $\mathcal{L}$. 
%%	Algorithm repeats this process until no more violating loop exists. 
%Finding this loop is easier than finding the shortest loop and it suffices to employ a BFS search from each node with cut-off  $1-$tol and reporting the first found back link that form a loop with less than cut-off. In our experience form most of the networks this process is in linear time.

Although simple, this algorithm is adequate for computing the modulus in the examples presented here,
on a Linux operating computer with Intel core i$7$ (and $2.80$ GHz base frequency) processor, for example.  More advanced parallel primal-dual algorithms are currently under development to treat modulus computations on larger networks.

\section{Clustering measure with modulus of family of loops}

Complex networks exhibit properties such as the small-world phenomenon \cite{watts1998collective}, scale-free degree distribution \cite{Barabasi99emergenceScaling}, and local clustering of nodes  \cite{watts1998collective}.
In social networks, when two individuals are acquainted it is probable that they have another friend in common, resulting in propeties of homophily for the network. For example, in friendship networks people introduce their friends to each other.
This transitivity property makes the real world networks different from synthetic random networks \cite{newman2003ego}.
However, this clustering tendency is difficult to quantify. 

A proposed measure of clustering for a node $v$ \cite{watts1998collective} is to compute the fraction of links between neighbors of $v$ that actually are in the network, over all possible ones.
The authors in \cite{caldarelli2004structure} pointed out the importance of closed paths (loops) in the cluster and discussed computation of the clustering coefficient using the density of loops with length $3$ (triangles).
Because this measure fails to describe the clustering of grid-like parts of the network, the authors improved the measure by counting quadrilaterals--loops with length $4$ or \emph{mutuality} in \cite{newman2003ego}--and proposed a new measure that considers different types of quadrilaterals.
Similarly, \cite{lind2005cycles} addresses bipartite networks, that lack triangles thus the standard clustering coefficient is not useful. In \cite{lind2005cycles}, \cite{lind2007new} and \cite{fronczak2002higher} the authors emphasize the importance of  longer loops in the network.
% Finally, \cite{latapy2008basic} proposed a different clustering coefficient for bipartite networks.
The authors in \cite{soffer2005network}, showed that clustering coefficient measures are highly correlated with degree, and they proposed a measure that preserves the degree sequence for the maximum possible links among neighbors of node $v$, thus avoiding correlation biases.
%\textbf{NEW:} 
Kim \textit{et al.} introduced \emph{local cycling coefficient} that quantifies local circle topologies by averaging the inverse length of loops passing the nodes \cite{kim2005cyclic}. They average this coefficient for all nodes to derive the degree of circulation in the network.

%None of the above methods considered weighted and directed networks. 
The authors in \cite{saramaki2007generalizations} introduce a version of clustering coefficient that considers weighted network, and \cite{opsahl2009clustering} propose a way to measure a general clustering coefficient for weighted and directed networks.

Numerous versions of clustering coefficients for different types of networks  expose the need for a generalized measure that works for a wide range of applications.
We apply the concept of modulus of families of loops as a tool to study structural properties of network clustering.
In this section we show that analysis of loops using modulus provides a general approach to the study of network clustering properties. We also propose a new clustering measure that can explain situations that conventional methods struggle to handle.

A network has a high clustering measure when most of the links are included in short loops that also visit nearby links.
 The standard method of counting triangles considers the smallest loops, while other methods consider the next shortest loops, i.e., quadrilaterals. A method must be devised to compare these loops and evaluate the combined influence to improve clustering measures \cite{newman2003ego}.
The previous section introduced a way to evaluate family of  loops using  modulus. Therefore, we propose a comprehensive modulus-based measure of clustering. 

The classical clustering coefficients that measure triangle density, are usually normalized by comparing the links in the networks (that form triangles) with all possible links between nodes, i.e., all possible triangles in the corresponding complete graph. Most real networks are far from being complete graphs (even locally), therefore, classical coefficients usually have small values, and they are correlated to the degree of the node \cite{soffer2005network}.

We normalize our clustering measure using the probabilistic interpretation in \eqref{eq:probInterp}. Modulus tries to spread expected usage as much as possible among the links of the network in order to minimize the expected overlap. However, the expected link usages are not always uniform. Define a uniform density $\rho_u(e)\equiv 1/3$ that is always admissible for loop modulus--because it penalizes all loops at least $1$. So its energy $\mathcal{E}_2(\rho_u)=|E|/9$
gives an upper bound for $\Mod_2(\mathcal{L})$.

\begin{figure}
\subfloat[]{%
  \includegraphics[clip,width=.5\columnwidth]{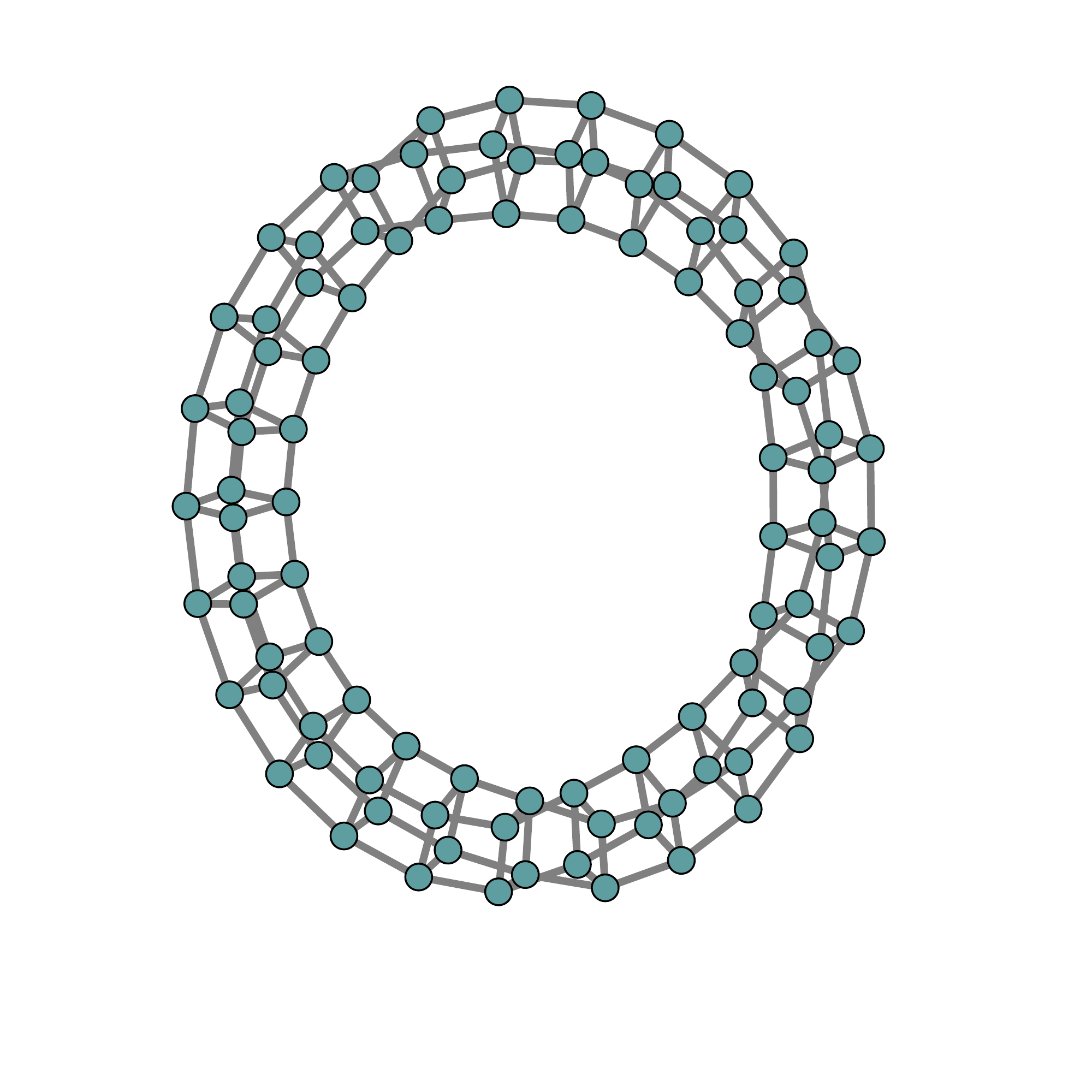}%
}
\subfloat[]{%
  \includegraphics[clip,width=.5\columnwidth]{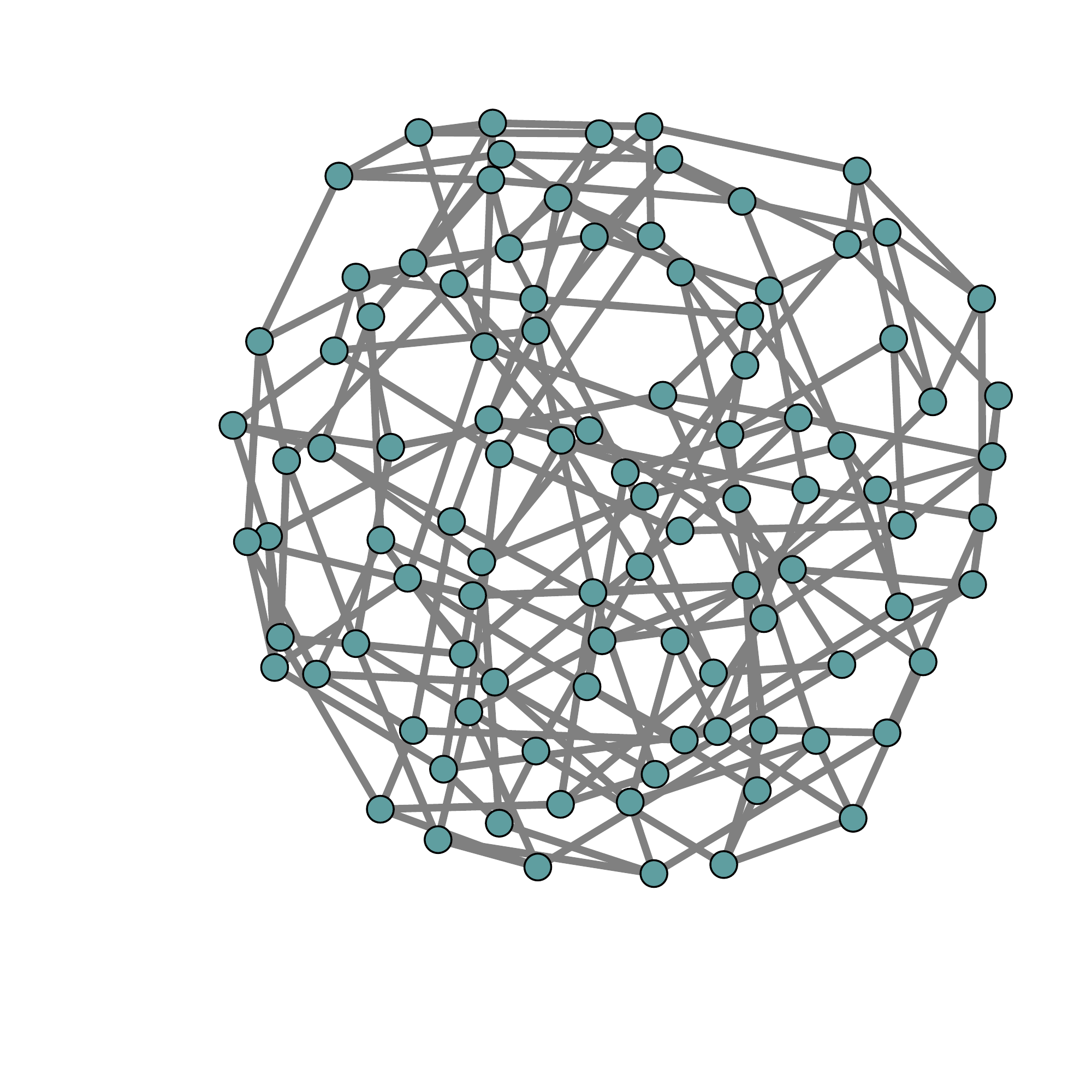}%
}
\caption{(a) A grid network with $ \deg= 4$ and $100$ nodes, (b) a random regular network with $\deg = 4$ and $100$ nodes. The proposed clustering measure is $C\left(\GG_{\text{grid}}\right)=56.25\%$, $C\left(\GG_\text{reg}\right) = 34\%$. Classical clustering coefficient gives zero for the grid and $2.4\%$ for the regular network and average square clustering coefficient is $14.7\%$ for the grid  and $0.4\%$ for the regular network. }\label{fig:GlobalCC}.
\end{figure}

Therefore, our proposed clustering measure takes the following form
\begin{equation}\label{eq:CC}
C_{\rm loop}(\GG) := \frac{9}{|E|}\Mod_2(\mathcal{L}),
\end{equation}
where $C_{\rm loop}$ is a measure of richness of actual link participation in important loops over the ideal case that all links participate equally in triangles.
For example, consider a grid as in Figure \ref{fig:GlobalCC}(a) with $100$ nodes and $200$ links. We compare its loop modulus with that of a random regular network with the same number of nodes and same degree as shown in Figure \ref{fig:GlobalCC}--these networks behave similar to the two extremes of small world networks \cite{watts1998collective}.  Since the classical methods use the number of triangles in a network, they give zero clustering coefficient to the grid and $2-3\%$ to the random regular network. The grid has square clustering coefficient $14.7\%$ and the random regular network square clustering is close to zero (we use square clustering introduced in \cite{lind2005cycles}). 
For each network in Figures \ref{fig:GlobalCC}(a) and \ref{fig:GlobalCC}(b):
$$\Mod_2\mathcal{L}_{\text{grid}} = 10.8\quad\text{ and }\quad \Mod_2\mathcal{L}_{\text{reg}} = 7.8.$$ Therefore, $C_{\rm loop}\left(\GG_{\text{grid}}\right)=54\%$
 which means the network is highly clustered and $C_{\rm loop}\left(\GG_\text{reg}\right) = 34\%$ is less clustered than grid.

\begin{figure}
\subfloat[]{%
  \includegraphics[clip,width=.5\columnwidth]{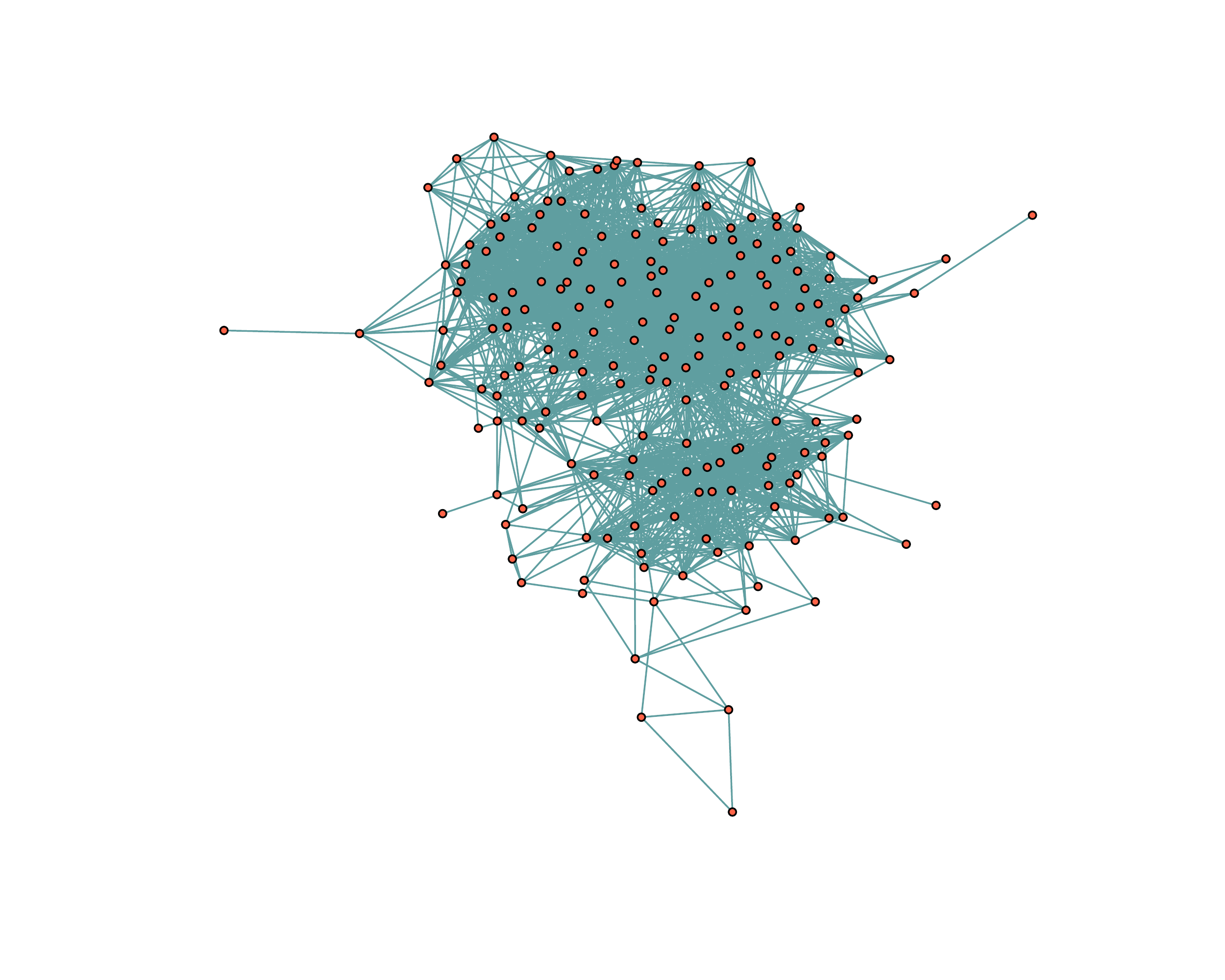}%
}
\subfloat[]{%
  \includegraphics[clip,width=.5\columnwidth]{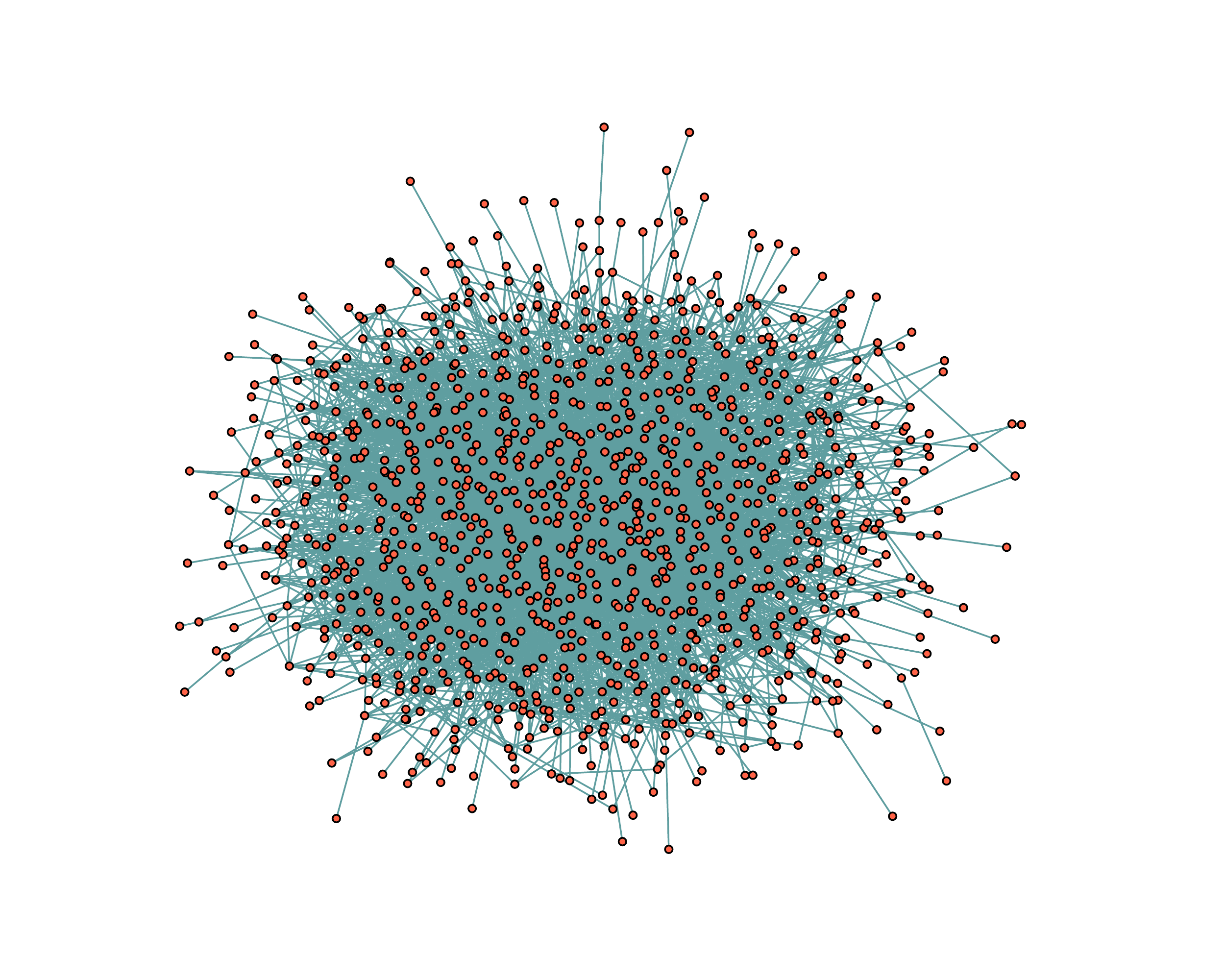}%
}

\subfloat[]{%
  \includegraphics[clip,width=.5\columnwidth]{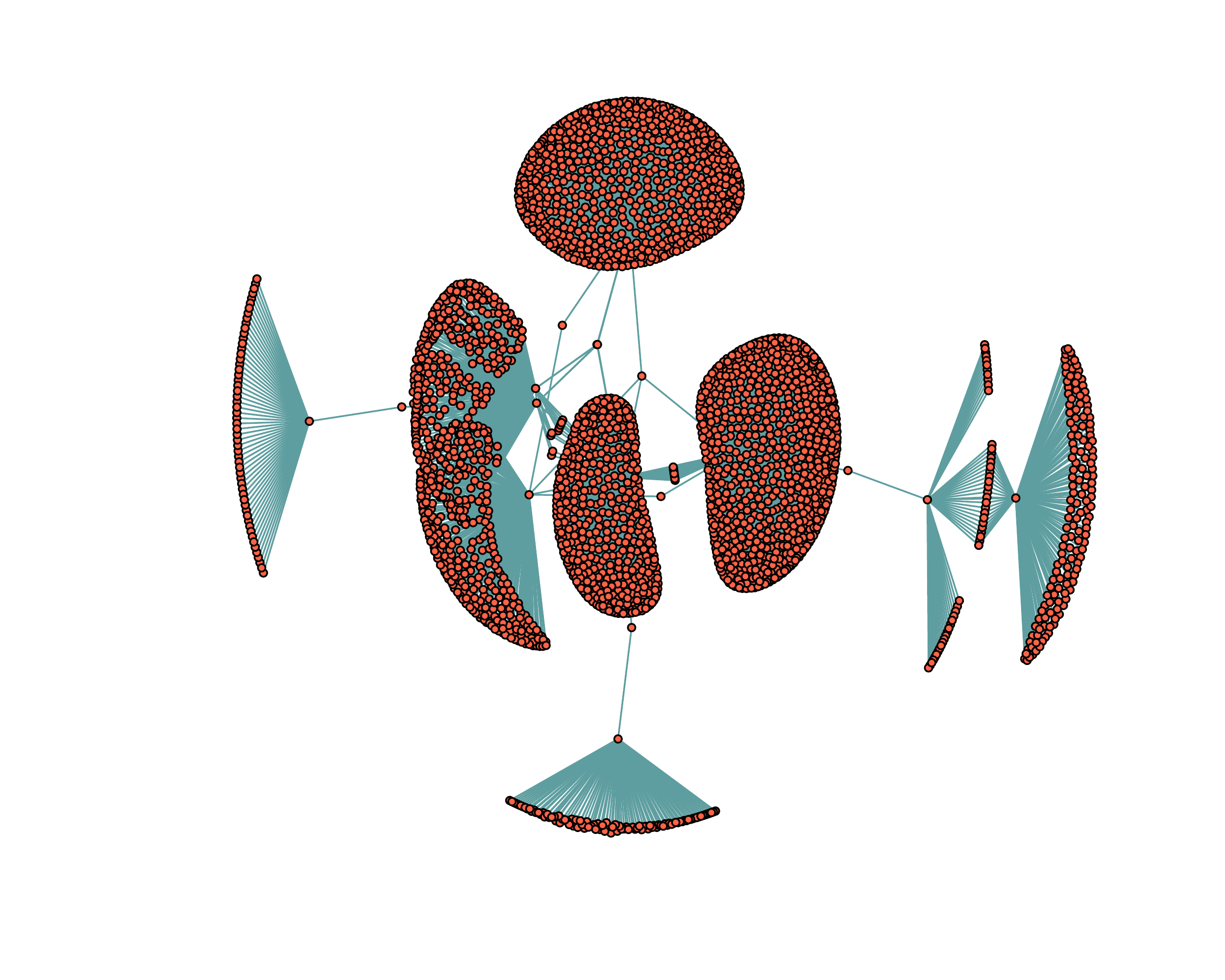}%
}
\subfloat[]{%
  \includegraphics[clip,width=.5\columnwidth]{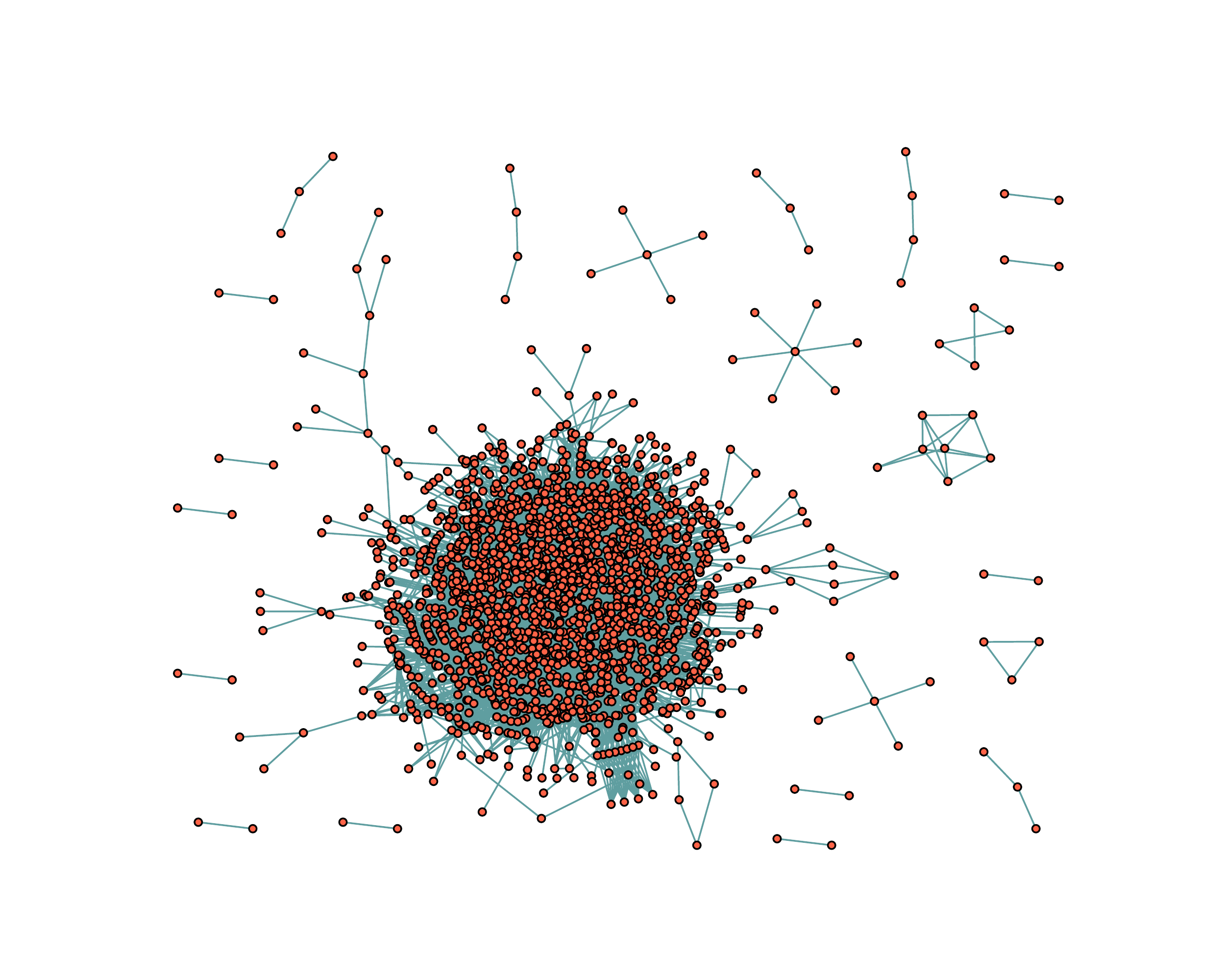}%
}
\caption{(a) Jazz musicians network  \cite{Jazz} with $C_{\rm loop} =10.0\%$; average triangle density $C =52.0\%$ and average square clustering $6.66\%$. (b) Email communication network in University Rovira i Virgili in Spain with $C_{\rm loop} =13.8\%$; average triangle density $C =16.6\%$  and average square clustering $1.46\%$ \cite{guimera2003self}. (c) An excerpt of Facebook network with $n = 2888$ and $m=2981$. Edges represent friendships between nodes \cite{mcauley2012learning} with $C_{\rm loop} = 3.7\%$; average triangle density $0.03\%$ and average square clustering $0.07\%$. (d) 
Friendship network of the website hamsterster.com \cite{Hamster}, with $n = 1858$ and $m=12534$. The clustering in the network is $C_{\rm loop} = 6.22\%$. The classical clustering coefficient (transitivity) is $9.04\%$ and average square clustering coefficient $6.78\%$.}\label{fig:JazzArenasFace}
\end{figure}

In some cases, our proposed measure gives different conclusions than the classical cluster coefficients. For example, let us compare the networks (a) and (b) in Figure \ref{fig:JazzArenasFace}. Network (a) is collaboration network between Jazz musicians \cite{Jazz} and network (b) is an email communication network at the University Rovira i Virgili in Spain  \cite{guimera2003self}. In the email communication network a very rich core is balanced by many stems on the periphery and the loop clustering measure is slightly higher than for the Jazz network. This goes in the opposite direction than the classical clustering coefficient result \cite{kunegis2014handbook}.  For the piece of the Facebook network in Figure \ref{fig:JazzArenasFace}(c)  \cite{mcauley2012learning}, the loop clustering value is slightly greater than the classical case, reflecting a certain amount of tightly knit communities. Finally, in the friendship network for the website \href{www.hamsterster.com}{hamsterster} \cite{Hamster}, the clustering measure and classical clustering coefficient give almost similar results.

Furthermore, we can isolate the contribution of triangles, squares, and higher order loops by considering  modulus of subfamilies of $\mathcal{L}$.
This can be done  assuming a hop-length cut-off for $\gamma$ in Algorithm \ref{alg:DS}. Moreover, the property of subadditivity (Property (e)) gives an upperbound for the aggregate effects.

\section{Weighting to enhance community detection algorithms}

Communities in networks are defined as groups of nodes that are closely knit together relative to the rest of the network. Real world networks, for example
social networks \cite{homans2013human} and
biological networks \cite{ravasz2002hierarchical}, comprise densely connected parts that are loosely connected with each other.
Finding these communities is crucial in analyzing the collective behavior of the network or in order to be able to make assumptions (meta population). These communities can be disjoint or overlapping. For a comprehensive review of the literature on this subject see \cite{fortunato2010community}.  

Radicchi et al. count the number of short loops that pass each link as a local measure for clustering \cite{radicchi2004defining}. To extend the method in \cite{radicchi2004defining} for low clustered networks, Vragovic et al. in \cite{vragovic2006network} consider general loops (with any length) passing the nodes to detect cluster nodes; although, compared to standard clustering methods, its results are not satisfying \cite{fortunato2010community}.

The authors in \cite{berry2011tolerating} define a new weighting for the network to improve modularity maximization methods for finding communities with sizes smaller than the resolution limit \cite{fortunato2007resolution}. The weigthing for a link comes from how many loops with length $3$ and $4$ it forms with the adjacent links. They show the effectiveness of their method on Lancichinetti, Fortunato, and Radicchi (LFR) benchmark networks. Also the authors in \cite{khadivi2011network} propose weighting the network  with a combination of link betweenness centrality \cite{freeman1977set} and their other measure {\it common neighbor ratio} to enhance community identification.
%Moeover, modulus can analyze loops in directed networks as well. 
Community detection in directed networks is a challenging problem  \cite{malliaros2013clustering}. \cite{klymko2014using} improved community detection in directed networks by weighting the network. They
consider seven different types of triangles and their respective contributions to the community structure. 

When a pair of nodes are in the same group it is more likely to have strong flow of communication among each other together with their groupmates and information tends to stay within communities. This emphasises the importance of having many non-overlapping short loops.

Analyzing loops in a network provides information about the cluster structure and emphasizes the importance of links in these clusters.
By (\ref{eq:probInterp}) the extremal density $\rho^*(e)$ measures the amount of important loops (see Section \ref{sec:probInterp}) passing through link $e$ (expected usage). Assuming members in the community shares a lot of cycles between themselves, thus $\rho^*(e)$ serves as a  measure of affinity for the nodes connected by $e$. In other words, nodes on  important  loops  are well connected to the rest of the group. 
In this section, we show that indeed preprocessing the network using $\rho^*(e)$  can improve network partitioning.

After we compute loop modulus  for a network, the extremal density $\rho^*(e)$ gives generic information about the structure of communities that contains many short loops and the importance of links in these clusters that generalize methods in \cite{radicchi2004defining} and \cite{vragovic2006network}. We can substantially  improve the performance of some partitioning methods such as spectral partitioning or modularity maximization heuristics by preprocessing the network into a weighted network with link weights $\rho^*(e)$. We can apply our methods to any weighted and directed network.

\begin{figure}
\subfloat[]{%
  \includegraphics[clip,width=.98\columnwidth]{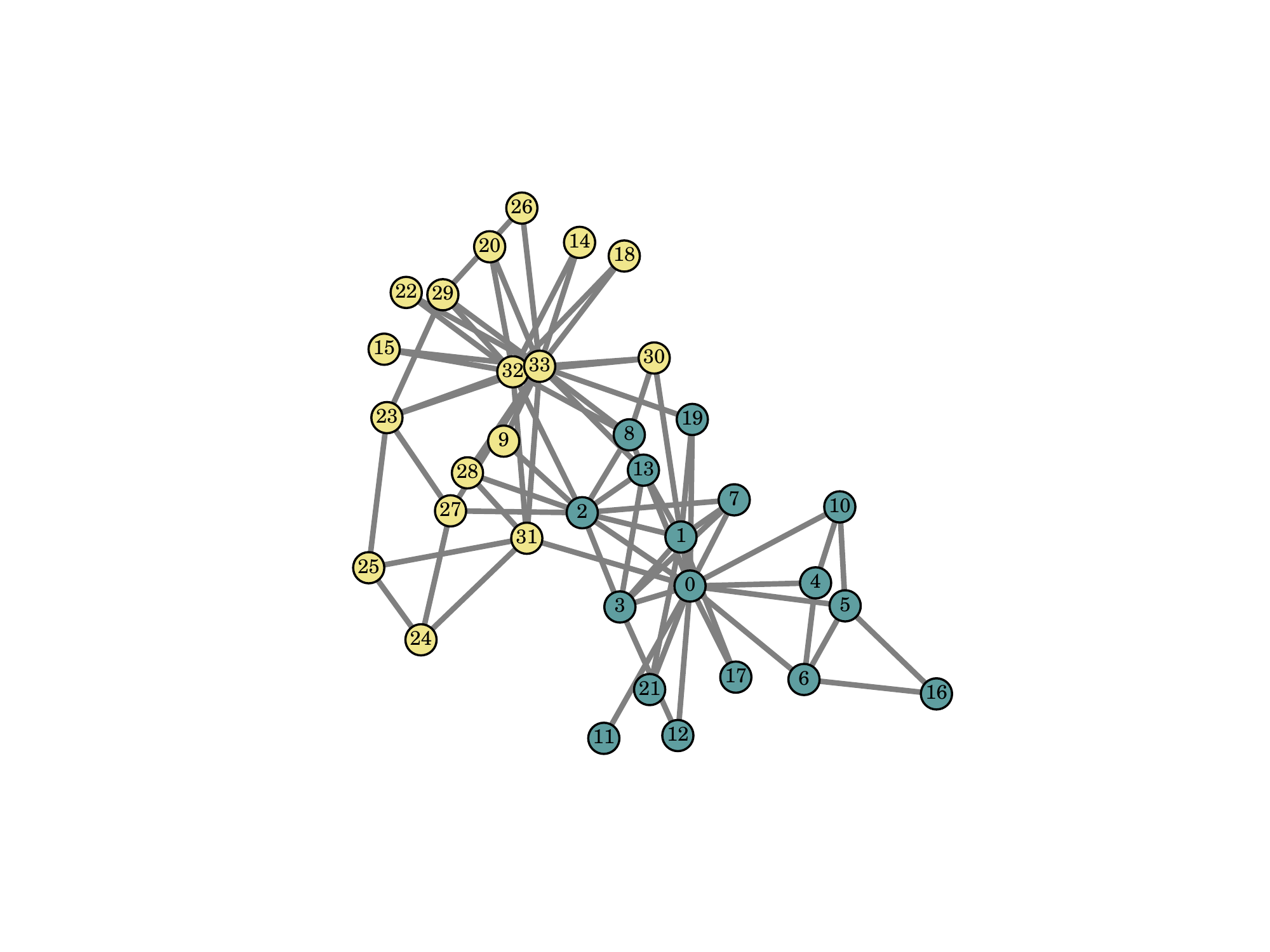}%
}\\
\subfloat[]{%
  \includegraphics[clip,width=.5\columnwidth]{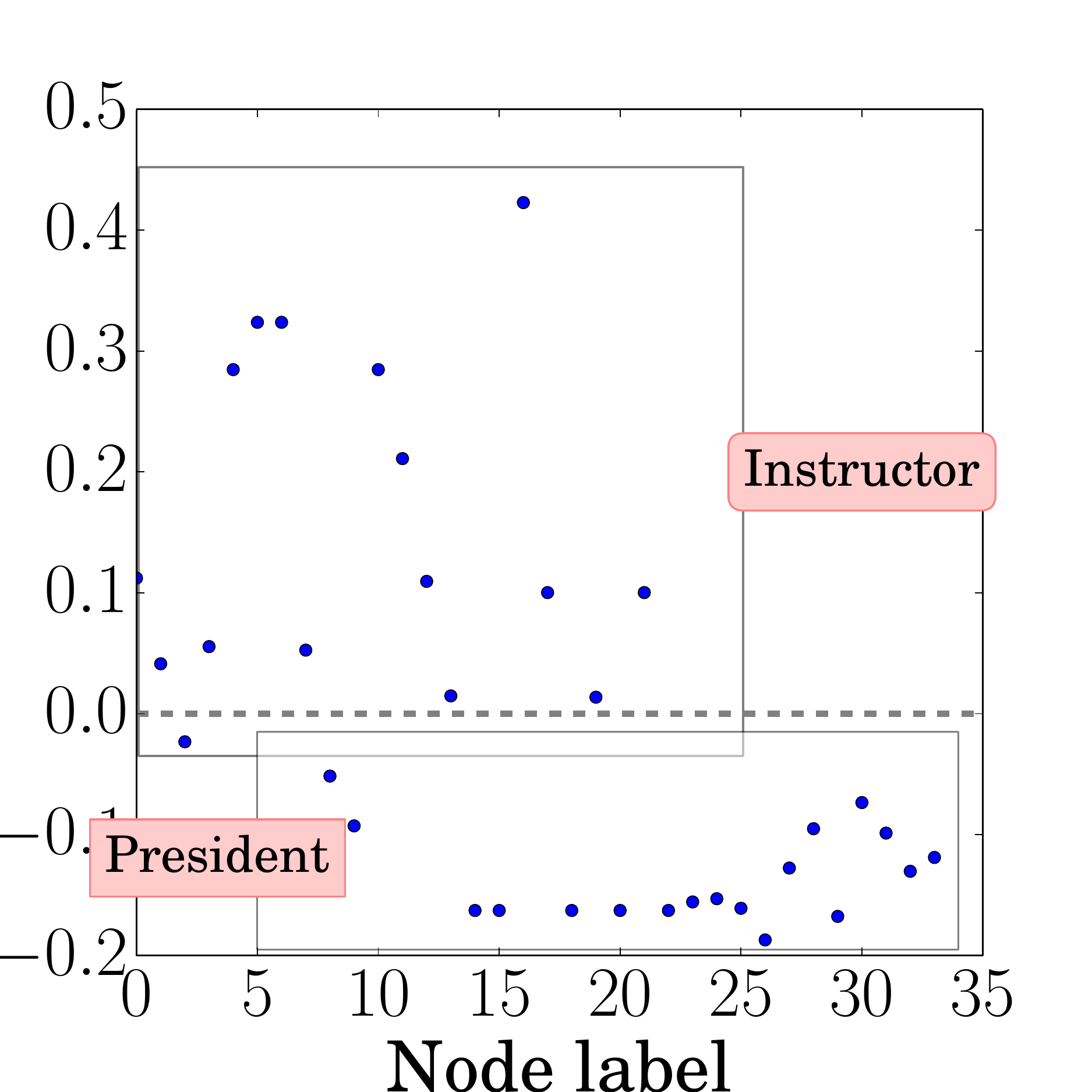}%
}
\subfloat[]{%
  \includegraphics[clip,width=.5\columnwidth]{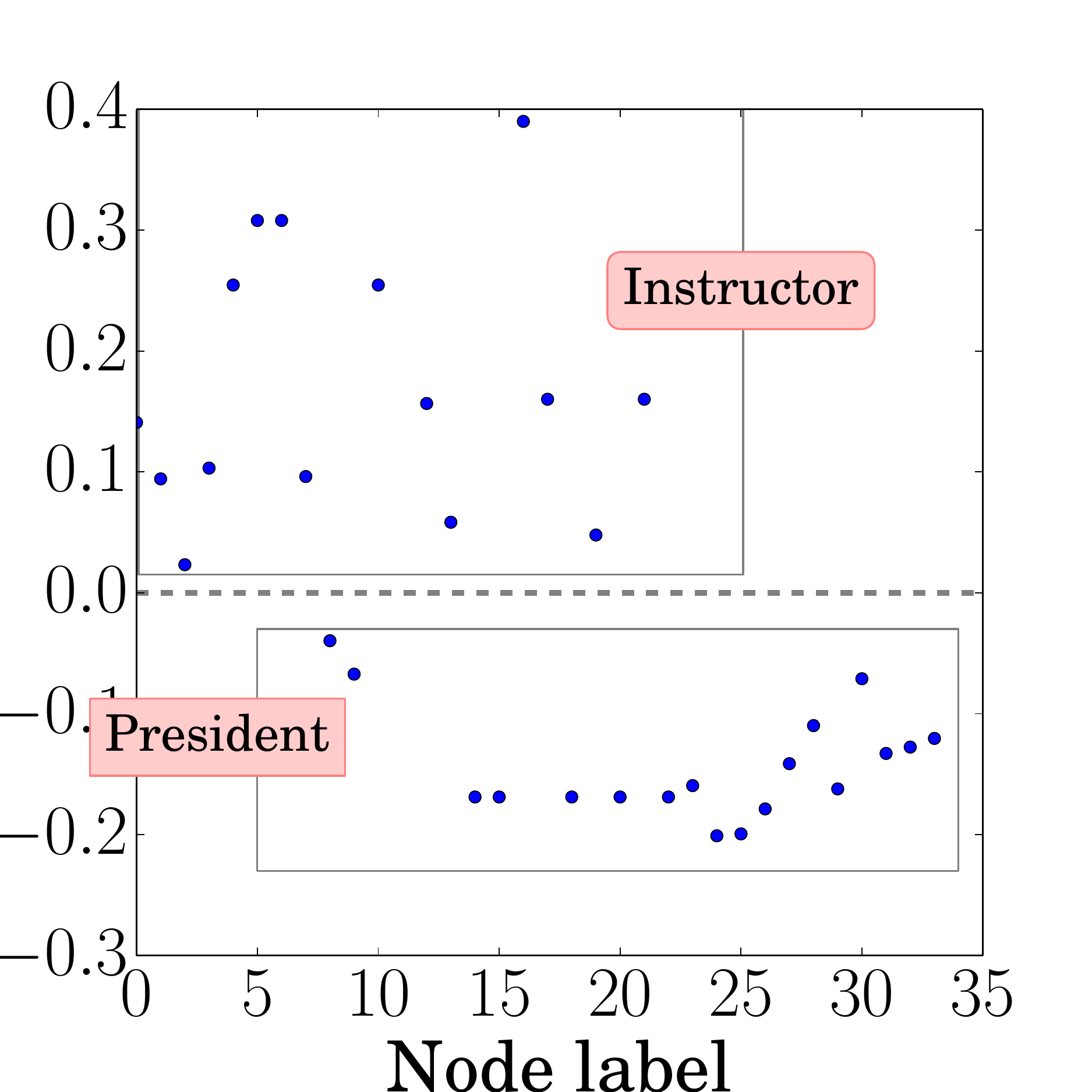}
}
\caption{(a) Zachary's karate club network \cite{ZacharyKarate} with the groups splitted after conflict. (b)-(c) Fiedler vector values corresponding with the node labels. (b) Spectral partitioning of Zachary's karate club network \cite{ZacharyKarate}, node $3$ is wrongly partitioned. (c) spectral partitioning of the same network weighted by Loop Modulus where nodes are correctly partitioned. }\label{fig:Karate}
\end{figure}

As the first example, we consider Zachary's Karate Club \cite{ZacharyKarate}--a friendships network at a university Karate club with $34$ members, see Figure \ref{fig:Karate}(a).
A conflict between the instructor and the club's president split the club into two groups. Finding the communities in this network is a basic benchmark test for partitioning algorithms \cite[Chapter 9]{barabasi2013network}.

To bisect this network, we use Fiedler vector bisection \cite{NewmanBook}
on both weighted and unweighted networks in Figures \ref{fig:Karate}(b) and (c). In the unweighted case, the bisection method failed to separate a node correctly and there are two nodes that are very close to the other cluster. Our weighting method does this clustering with complete accuracy.

\begin{figure}
\subfloat[]{%
  \includegraphics[clip,width=.8\columnwidth]{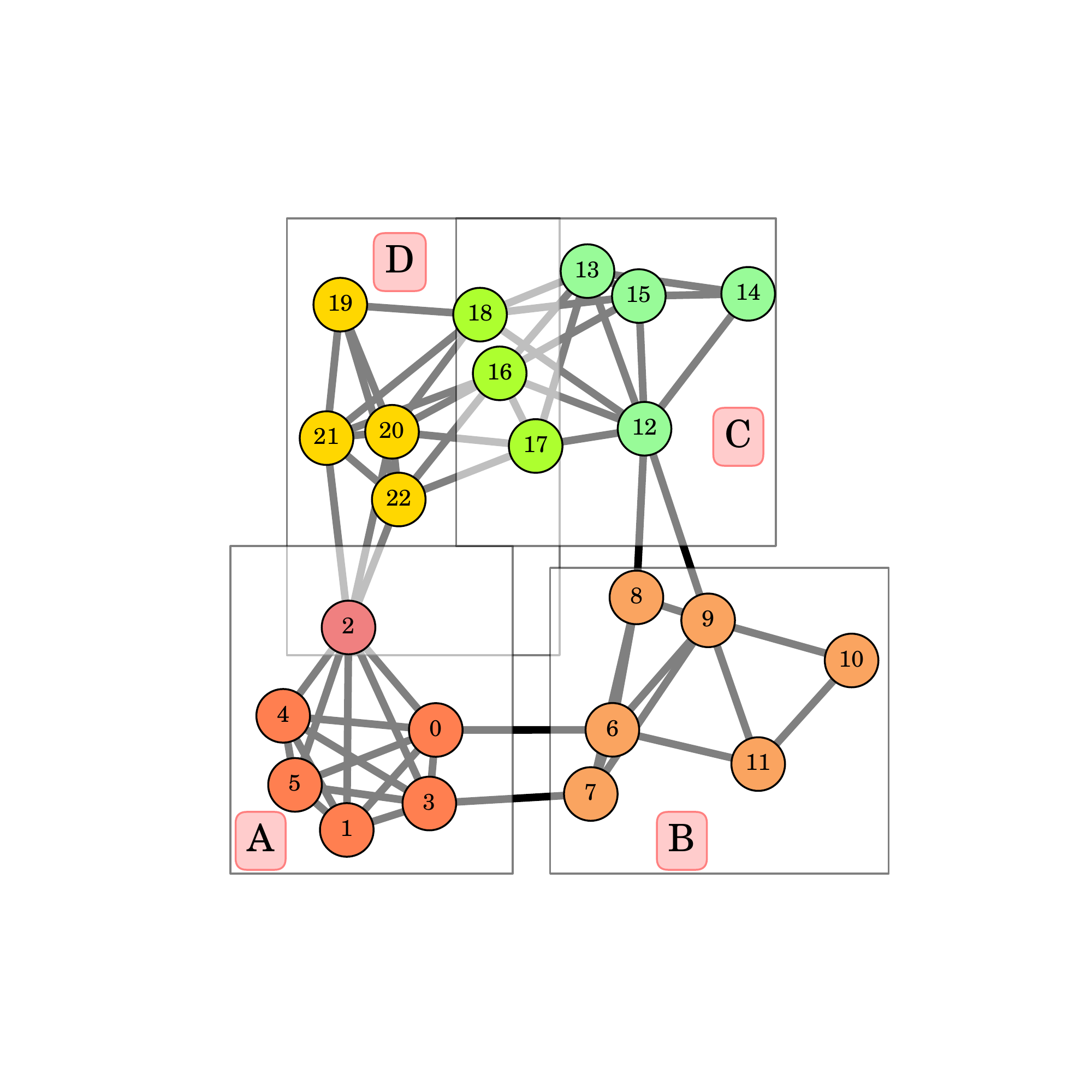}%
}\\
\subfloat[]{%
  \includegraphics[clip,width=.5\columnwidth]{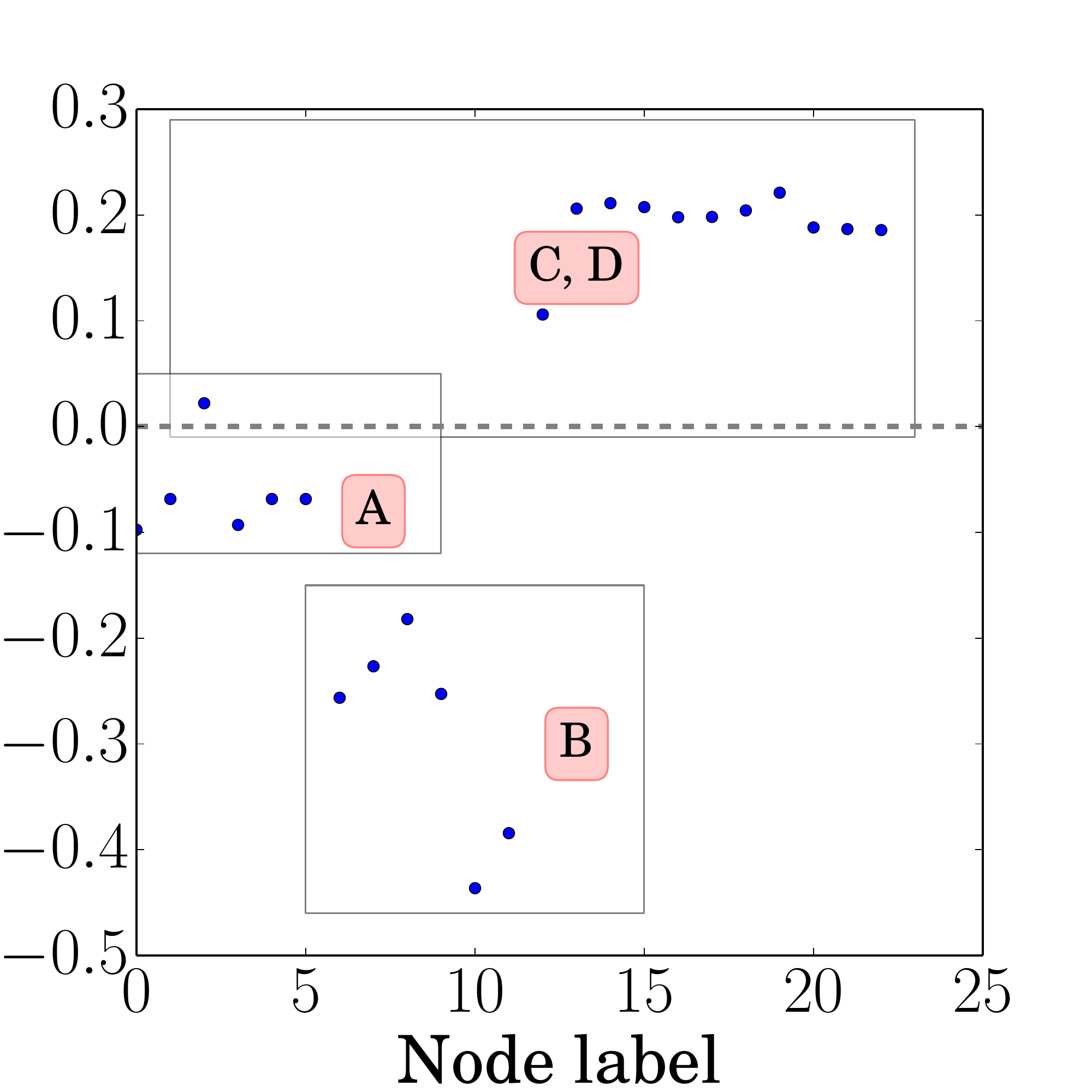}%
}
\subfloat[]{%
  \includegraphics[clip,width=.5\columnwidth]{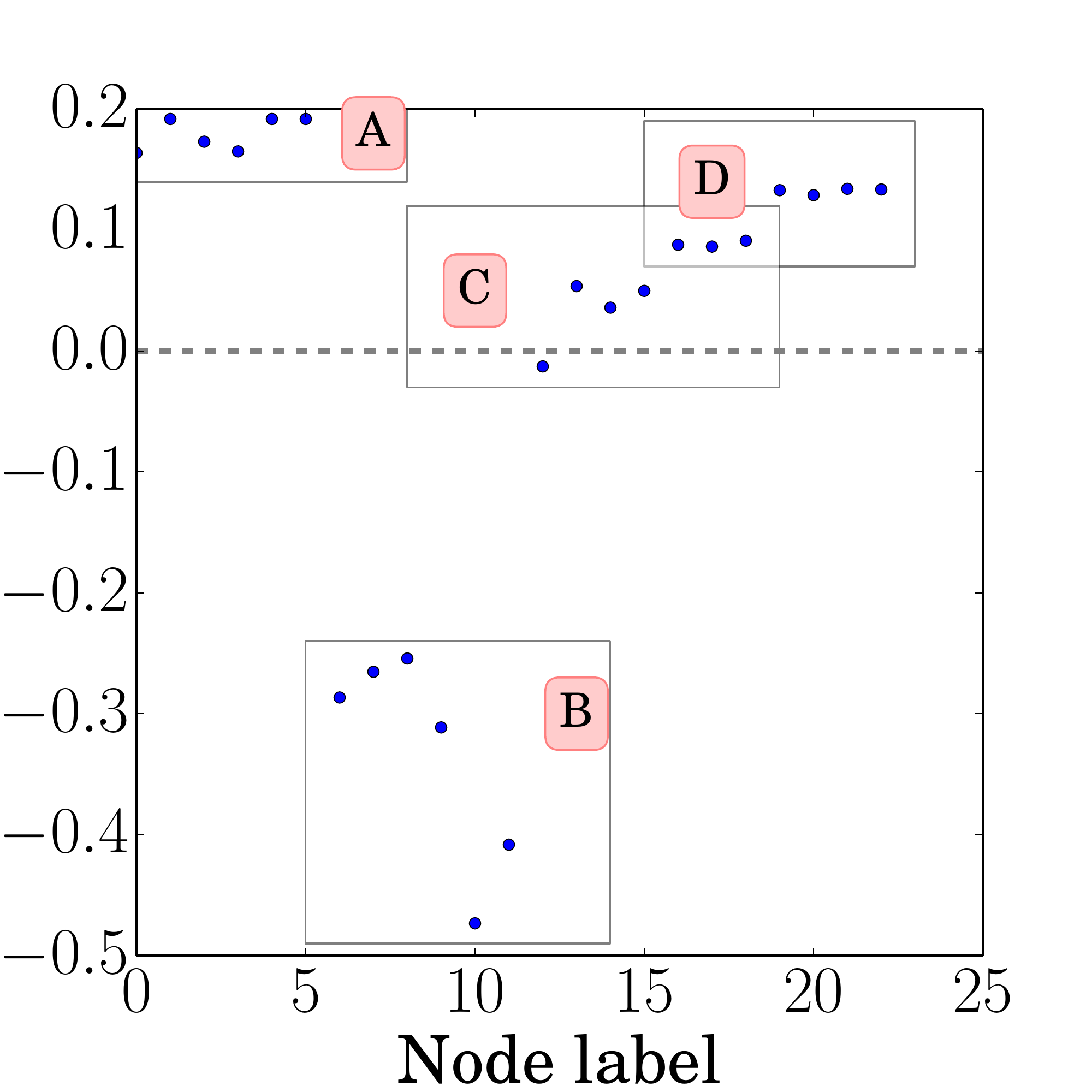}%
}
\caption{(a) A network partitioned by Palla et. al. \cite{Palla}. Nodes $16$, $17$ and $18$ are shared between C and D groups and Node $2$ is shared between D and A groups. (b) Fiedler vector of the network, (c) Fiedler vector of the weighted network  by Loop Modulus where overlapping groups can be distinguished.}\label{fig:Overlap}
\end{figure}

It may be useful to allow for overlapping communities. For instance,
a node can be a member of different communities, such as family, sport club, workplace, etc \cite{palla2007quantifying}.
Although bisection methods alone are unable to detect overlapping communities, we see that loop modulus can augment these methods by distinguishing nested partitions in networks with overlapping communities in the next example. Figures \ref{fig:Overlap} (a)--(c) show a network that is partitioned by Palla et al. \cite{Palla}. We compute the Fiedler vector in both unweighted and weighted cases. As shown, the unweighted method failed to separate C and D overlapping communities, while the  weighted method does distinguish them with the overlapping part.

To show the effectiveness of the weighting method in a more standard fashion, we consider two popular heuristics for modularity maximization; greedy modularity optimization method by Clauset, Newman, and Moore (CNM) \cite{clauset2004finding} and the Louvian method \cite{blondel2008fast} on the LFR benchmarks \cite{lancichinetti2008benchmark}. The LFR benchmarks allow the user to specify the community size distribution along with the degree distribution, offering more realistic benchmarks  than the Girvan-Newman benchmarks \cite{girvan2002community}. 
%In \cite[Chapter 6]{barabasi2013network}, the author discusses the effectiveness of different community identification methods on LFR benchmarks and how CNM method performs poorly on them.
% due to resolution limit \cite{fortunato2007resolution}. 
We show re-weighting the network, using $\rho^*(e)$ from loop modulus, improve both CNM and Louvian substantially.

\begin{figure}
\subfloat[]{%
  \includegraphics[clip,width=.3\columnwidth]{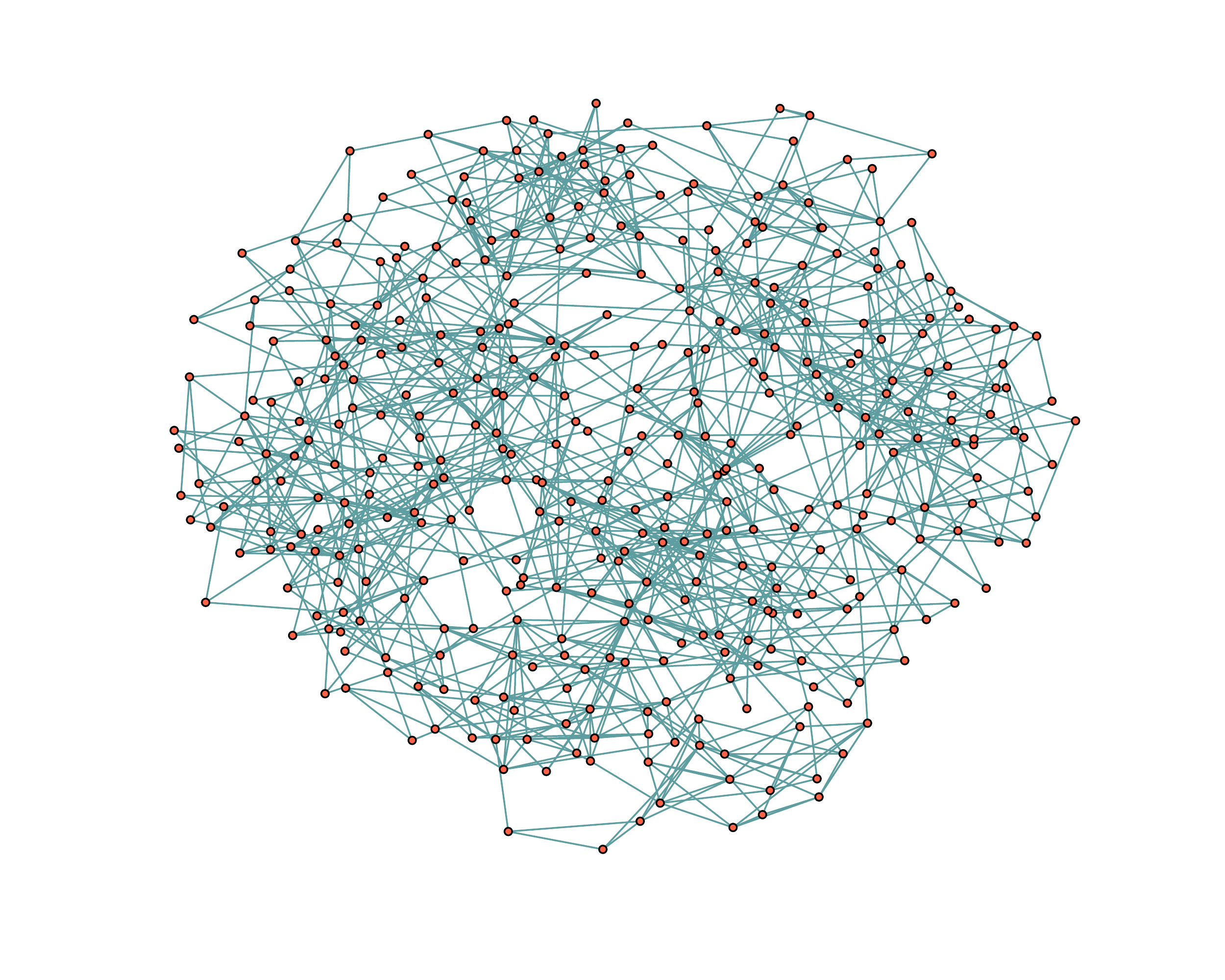}%
}
\subfloat[]{%
  \includegraphics[clip,width=.3\columnwidth]{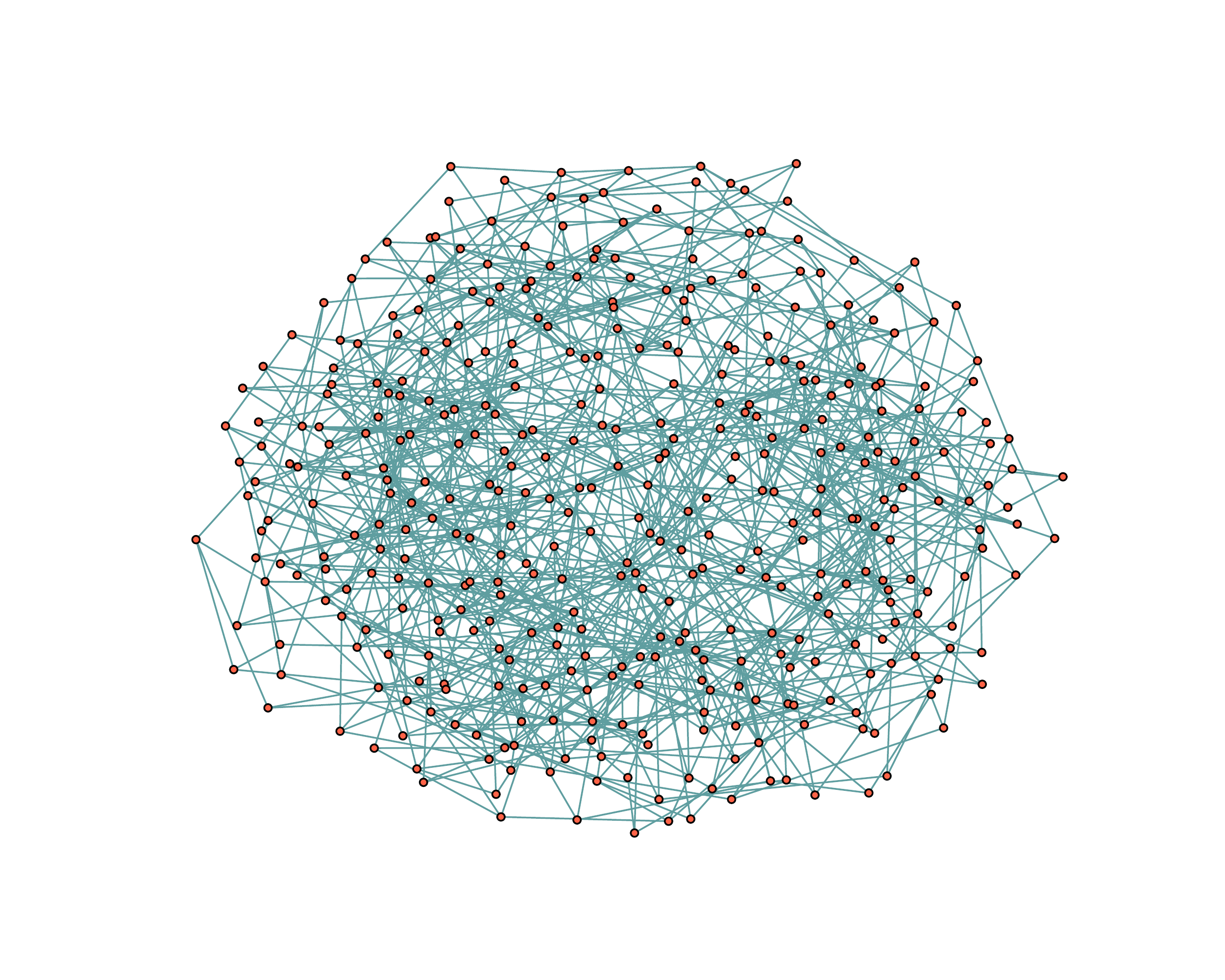}%
}
\subfloat[]{%
  \includegraphics[clip,width=.3\columnwidth]{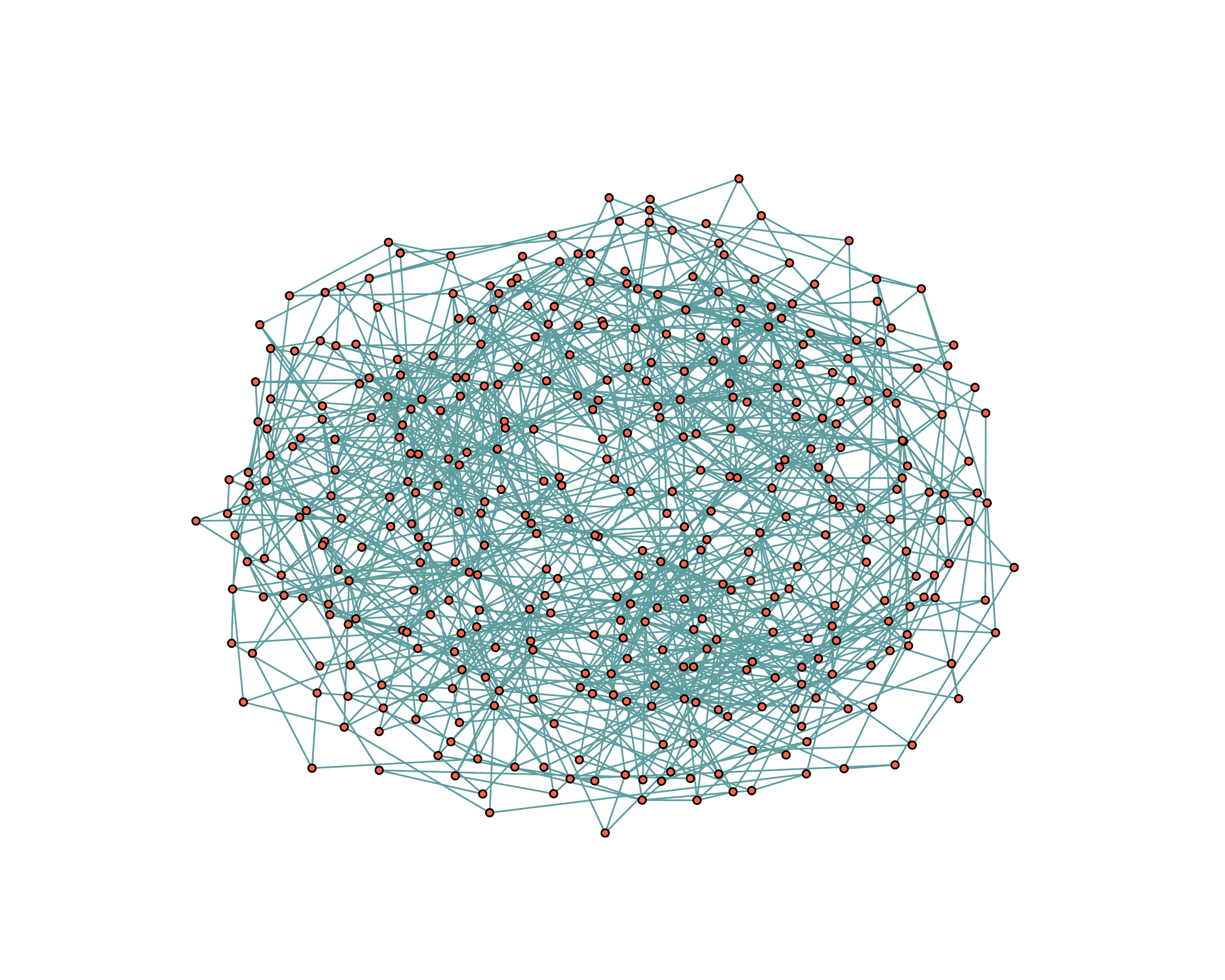}%
}

\subfloat[]{%
  \includegraphics[clip,width=.9\columnwidth]{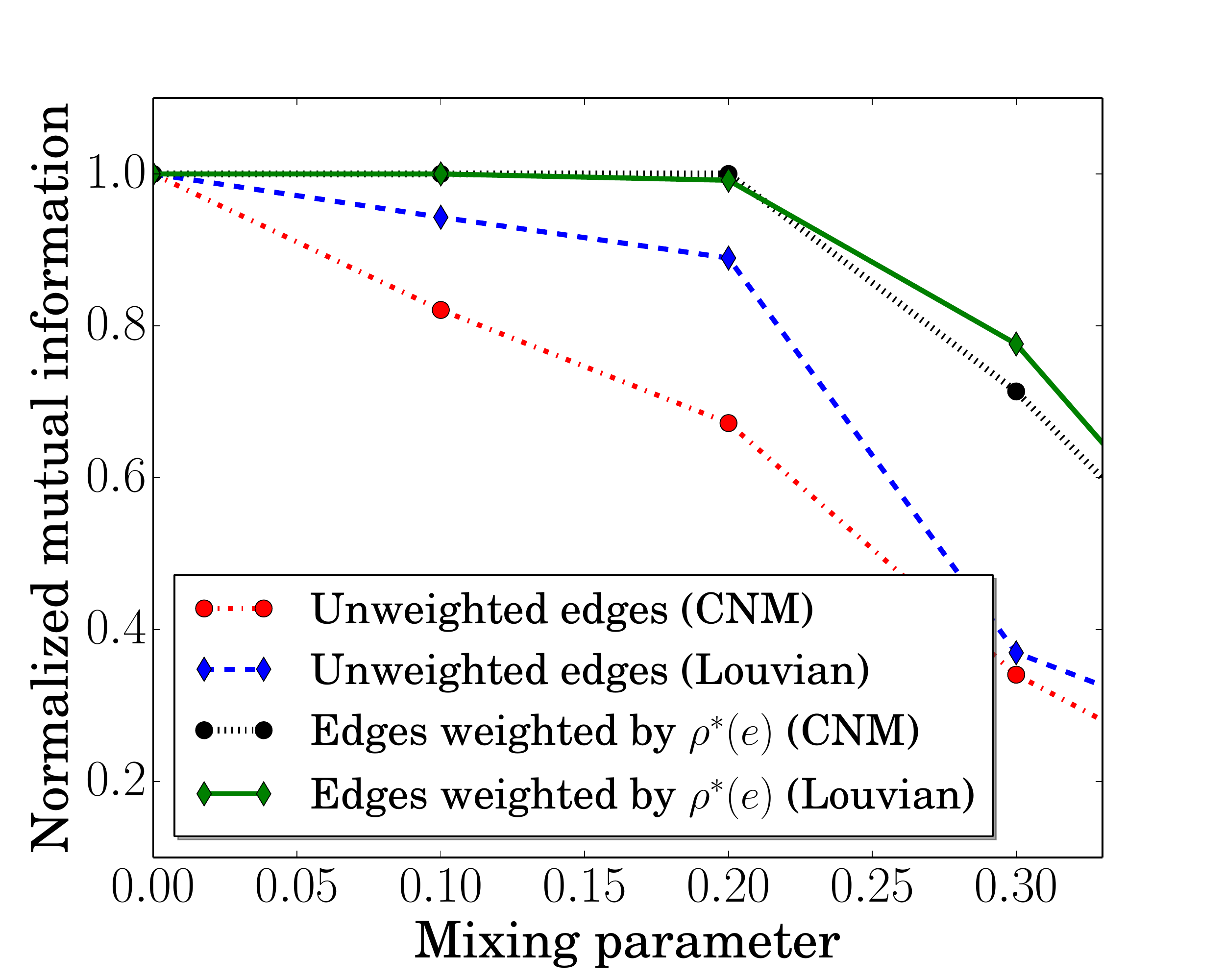}%
}
\caption{(a)-(c) Networks are produced by LFR benchmark with size $400$ nodes, mean degree $5$, maximum degree $10$,  and community sizes ranging from $20-40$.
The mixing rate $\mu$, for adjusing ratio of intra-communities links over all links are $0.1$, $0.2$, and $0.3$.
(d) The
plot depicts the normalized mutual information for community memberships found by Greedy modularity optimization (CNM) and Louvian method. Both the CNW and Louvian methods perform a better task on re-weighted networks.}\label{fig:InfoMixLFR}
\end{figure}

In Figure \ref{fig:InfoMixLFR}(a)-(c), three networks are produced by the LFR benchmark with $400$ nodes, mean degree $5$, maximum degree $10$,  and community sizes ranging from $20-40$ nodes.
The interconnectedness of various communities is measured by the mixing rate $\mu$.  We plot the mutual information \cite{danon2005comparing} for both the derived membership from CNM and Louvian on each network and the weighted version and compare them to the ground truth from LFR in Figure \ref{fig:InfoMixLFR}. As we observed, both the CNW and Louvian algorithms perform better on re-weighted networks using modulus. 
%This can be partially explained, since in the unweighted  case the CNM algorithm  fails to capture small clusters below the resolution limit (see \cite{barabasi2013network}).

\section{Conclusion}
In this paper, we use modulus of family of loops to analyze loop structures in networks. We showed that loop modulus quantifies the richness of loops in the network and we used it to measure clustering. The extremal densities found for loop modulus represent the probability of link participation in important loops.  We showed that performance of community detection methods such as spectral bisection and modularity maximization partitioning can be improved by weighting networks with their extremal densities derived from loop modulus. 
Although, we present some applications of loop modulus,  analyzing loop structures on the network can expose different aspects of the network, such as various dynamics on the network, e.g., synchronization and propagation \cite{li2010consensus, kuramoto1975self,van2011n} as well as analyzing complexity of networks \cite{butts2000axiomatic}. 
\section{ACKNOWLEDGMENTS}
The authors are thankful to anonymous reviewer's for their insightful comments that improved the initial manuscript and provide directions for future work.
Thanks also to Michael Higgins and his research group for their valuable suggestions.
This work is funded by the National Science Foundation under Grant No. DMS-1515810.

%%---------------========================-------------------===========
%\section{Conclusion}

\bibliographystyle{apsrev4-1}
\bibliography{refs}

\begin{thebibliography}{56}%
\makeatletter
\providecommand \@ifxundefined [1]{%
 \@ifx{#1\undefined}
}%
\providecommand \@ifnum [1]{%
 \ifnum #1\expandafter \@firstoftwo
 \else \expandafter \@secondoftwo
 \fi
}%
\providecommand \@ifx [1]{%
 \ifx #1\expandafter \@firstoftwo
 \else \expandafter \@secondoftwo
 \fi
}%
\providecommand \natexlab [1]{#1}%
\providecommand \enquote  [1]{``#1''}%
\providecommand \bibnamefont  [1]{#1}%
\providecommand \bibfnamefont [1]{#1}%
\providecommand \citenamefont [1]{#1}%
\providecommand \href@noop [0]{\@secondoftwo}%
\providecommand \href [0]{\begingroup \@sanitize@url \@href}%
\providecommand \@href[1]{\@@startlink{#1}\@@href}%
\providecommand \@@href[1]{\endgroup#1\@@endlink}%
\providecommand \@sanitize@url [0]{\catcode `\\12\catcode `\$12\catcode
  `\&12\catcode `\#12\catcode `\^12\catcode `\_12\catcode `\%12\relax}%
\providecommand \@@startlink[1]{}%
\providecommand \@@endlink[0]{}%
\providecommand \url  [0]{\begingroup\@sanitize@url \@url }%
\providecommand \@url [1]{\endgroup\@href {#1}{\urlprefix }}%
\providecommand \urlprefix  [0]{URL }%
\providecommand \Eprint [0]{\href }%
\providecommand \doibase [0]{http://dx.doi.org/}%
\providecommand \selectlanguage [0]{\@gobble}%
\providecommand \bibinfo  [0]{\@secondoftwo}%
\providecommand \bibfield  [0]{\@secondoftwo}%
\providecommand \translation [1]{[#1]}%
\providecommand \BibitemOpen [0]{}%
\providecommand \bibitemStop [0]{}%
\providecommand \bibitemNoStop [0]{.\EOS\space}%
\providecommand \EOS [0]{\spacefactor3000\relax}%
\providecommand \BibitemShut  [1]{\csname bibitem#1\endcsname}%
\let\auto@bib@innerbib\@empty
%</preamble>
\bibitem [{\citenamefont {Milo}\ \emph {et~al.}(2002)\citenamefont {Milo},
  \citenamefont {Shen-Orr}, \citenamefont {Itzkovitz}, \citenamefont {Kashtan},
  \citenamefont {Chklovskii},\ and\ \citenamefont {Alon}}]{milo2002network}%
  \BibitemOpen
  \bibfield  {author} {\bibinfo {author} {\bibfnamefont {R.}~\bibnamefont
  {Milo}}, \bibinfo {author} {\bibfnamefont {S.}~\bibnamefont {Shen-Orr}},
  \bibinfo {author} {\bibfnamefont {S.}~\bibnamefont {Itzkovitz}}, \bibinfo
  {author} {\bibfnamefont {N.}~\bibnamefont {Kashtan}}, \bibinfo {author}
  {\bibfnamefont {D.}~\bibnamefont {Chklovskii}}, \ and\ \bibinfo {author}
  {\bibfnamefont {U.}~\bibnamefont {Alon}},\ }\href@noop {} {\bibfield
  {journal} {\bibinfo  {journal} {Science}\ }\textbf {\bibinfo {volume}
  {298}},\ \bibinfo {pages} {824} (\bibinfo {year} {2002})}\BibitemShut
  {NoStop}%
\bibitem [{\citenamefont {Mugisha}\ and\ \citenamefont
  {Zhou}(2016)}]{mugisha2016identifying}%
  \BibitemOpen
  \bibfield  {author} {\bibinfo {author} {\bibfnamefont {S.}~\bibnamefont
  {Mugisha}}\ and\ \bibinfo {author} {\bibfnamefont {H.-J.}\ \bibnamefont
  {Zhou}},\ }\href@noop {} {\bibfield  {journal} {\bibinfo  {journal} {arXiv
  preprint arXiv:1603.05781}\ } (\bibinfo {year} {2016})}\BibitemShut {NoStop}%
\bibitem [{\citenamefont {Petermann}\ and\ \citenamefont
  {De~Los~Rios}(2004)}]{petermann2004role}%
  \BibitemOpen
  \bibfield  {author} {\bibinfo {author} {\bibfnamefont {T.}~\bibnamefont
  {Petermann}}\ and\ \bibinfo {author} {\bibfnamefont {P.}~\bibnamefont
  {De~Los~Rios}},\ }\href@noop {} {\bibfield  {journal} {\bibinfo  {journal}
  {Physical Review E}\ }\textbf {\bibinfo {volume} {69}},\ \bibinfo {pages}
  {066116} (\bibinfo {year} {2004})}\BibitemShut {NoStop}%
\bibitem [{\citenamefont {Newman}(2003{\natexlab{a}})}]{newman2003structure}%
  \BibitemOpen
  \bibfield  {author} {\bibinfo {author} {\bibfnamefont {M.~E.}\ \bibnamefont
  {Newman}},\ }\href@noop {} {\bibfield  {journal} {\bibinfo  {journal} {SIAM
  review}\ }\textbf {\bibinfo {volume} {45}},\ \bibinfo {pages} {167} (\bibinfo
  {year} {2003}{\natexlab{a}})}\BibitemShut {NoStop}%
\bibitem [{\citenamefont {Lind}\ \emph {et~al.}(2005)\citenamefont {Lind},
  \citenamefont {Gonz{\'a}lez},\ and\ \citenamefont
  {Herrmann}}]{lind2005cycles}%
  \BibitemOpen
  \bibfield  {author} {\bibinfo {author} {\bibfnamefont {P.~G.}\ \bibnamefont
  {Lind}}, \bibinfo {author} {\bibfnamefont {M.~C.}\ \bibnamefont
  {Gonz{\'a}lez}}, \ and\ \bibinfo {author} {\bibfnamefont {H.~J.}\
  \bibnamefont {Herrmann}},\ }\href@noop {} {\bibfield  {journal} {\bibinfo
  {journal} {Physical review E}\ }\textbf {\bibinfo {volume} {72}},\ \bibinfo
  {pages} {056127} (\bibinfo {year} {2005})}\BibitemShut {NoStop}%
\bibitem [{\citenamefont {Bianconi}\ and\ \citenamefont
  {Capocci}(2003)}]{bianconi2003number}%
  \BibitemOpen
  \bibfield  {author} {\bibinfo {author} {\bibfnamefont {G.}~\bibnamefont
  {Bianconi}}\ and\ \bibinfo {author} {\bibfnamefont {A.}~\bibnamefont
  {Capocci}},\ }\href@noop {} {\bibfield  {journal} {\bibinfo  {journal}
  {Physical review letters}\ }\textbf {\bibinfo {volume} {90}},\ \bibinfo
  {pages} {078701} (\bibinfo {year} {2003})}\BibitemShut {NoStop}%
\bibitem [{\citenamefont {Kim}\ and\ \citenamefont
  {Kim}(2005)}]{kim2005cyclic}%
  \BibitemOpen
  \bibfield  {author} {\bibinfo {author} {\bibfnamefont {H.-J.}\ \bibnamefont
  {Kim}}\ and\ \bibinfo {author} {\bibfnamefont {J.~M.}\ \bibnamefont {Kim}},\
  }\href@noop {} {\bibfield  {journal} {\bibinfo  {journal} {Physical Review
  E}\ }\textbf {\bibinfo {volume} {72}},\ \bibinfo {pages} {036109} (\bibinfo
  {year} {2005})}\BibitemShut {NoStop}%
\bibitem [{\citenamefont {Albin}\ \emph {et~al.}(pear)\citenamefont {Albin},
  \citenamefont {Poggi-Corradini}, \citenamefont {Darabi~Sahneh},\ and\
  \citenamefont {Goering}}]{Modulus}%
  \BibitemOpen
  \bibfield  {author} {\bibinfo {author} {\bibfnamefont {N.}~\bibnamefont
  {Albin}}, \bibinfo {author} {\bibfnamefont {P.}~\bibnamefont
  {Poggi-Corradini}}, \bibinfo {author} {\bibfnamefont {F.}~\bibnamefont
  {Darabi~Sahneh}}, \ and\ \bibinfo {author} {\bibfnamefont {M.}~\bibnamefont
  {Goering}},\ }in\ \href@noop {} {\emph {\bibinfo {booktitle} {Proceedings of
  Complex Analysis and Dynamical Systems VII}}}\ (\bibinfo {year} {to appear})\
  \bibinfo {note} {http://arxiv.org/abs/1401.7640}\BibitemShut {NoStop}%
\bibitem [{\citenamefont {Albin}\ \emph {et~al.}()\citenamefont {Albin},
  \citenamefont {Brunner}, \citenamefont {Perez}, \citenamefont
  {Poggi-Corradini},\ and\ \citenamefont {Wiens}}]{Albin2014}%
  \BibitemOpen
  \bibfield  {author} {\bibinfo {author} {\bibfnamefont {N.}~\bibnamefont
  {Albin}}, \bibinfo {author} {\bibfnamefont {M.}~\bibnamefont {Brunner}},
  \bibinfo {author} {\bibfnamefont {R.}~\bibnamefont {Perez}}, \bibinfo
  {author} {\bibfnamefont {P.}~\bibnamefont {Poggi-Corradini}}, \ and\ \bibinfo
  {author} {\bibfnamefont {N.}~\bibnamefont {Wiens}},\ }\href@noop {} {\
  }\bibinfo {note} {Http://arxiv.org/abs/1504.02418}\BibitemShut {NoStop}%
\bibitem [{\citenamefont {Albin}\ and\ \citenamefont
  {Poggi-Corradini}(2016)}]{albin2016minimal}%
  \BibitemOpen
  \bibfield  {author} {\bibinfo {author} {\bibfnamefont {N.}~\bibnamefont
  {Albin}}\ and\ \bibinfo {author} {\bibfnamefont {P.}~\bibnamefont
  {Poggi-Corradini}},\ }\href@noop {} {\bibfield  {journal} {\bibinfo
  {journal} {arXiv preprint arXiv:1605.08462}\ } (\bibinfo {year}
  {2016})}\BibitemShut {NoStop}%
\bibitem [{\citenamefont {Ahlfors}(1973)}]{ahlfors1973}%
  \BibitemOpen
  \bibfield  {author} {\bibinfo {author} {\bibfnamefont {L.~V.}\ \bibnamefont
  {Ahlfors}},\ }\href@noop {} {\emph {\bibinfo {title} {Conformal invariants:
  topics in geometric function theory}}}\ (\bibinfo  {publisher} {McGraw-Hill
  Book Co.},\ \bibinfo {address} {New York},\ \bibinfo {year} {1973})\ pp.\
  \bibinfo {pages} {ix+157},\ \bibinfo {note} {mcGraw-Hill Series in Higher
  Mathematics}\BibitemShut {NoStop}%
\bibitem [{\citenamefont {Duffin}(1962)}]{duffin:1962jmaa}%
  \BibitemOpen
  \bibfield  {author} {\bibinfo {author} {\bibfnamefont {R.~J.}\ \bibnamefont
  {Duffin}},\ }\href@noop {} {\bibfield  {journal} {\bibinfo  {journal} {J.
  Math. Anal. Appl.}\ }\textbf {\bibinfo {volume} {5}},\ \bibinfo {pages} {200}
  (\bibinfo {year} {1962})}\BibitemShut {NoStop}%
\bibitem [{\citenamefont {Schramm}(1993)}]{schramm:1993israel}%
  \BibitemOpen
  \bibfield  {author} {\bibinfo {author} {\bibfnamefont {O.}~\bibnamefont
  {Schramm}},\ }\href@noop {} {\bibfield  {journal} {\bibinfo  {journal}
  {Israel Journal of Mathematics}\ }\textbf {\bibinfo {volume} {84}},\ \bibinfo
  {pages} {97} (\bibinfo {year} {1993})}\BibitemShut {NoStop}%
\bibitem [{\citenamefont {Shakeri}\ \emph {et~al.}(2016)\citenamefont
  {Shakeri}, \citenamefont {Poggi-Corradini}, \citenamefont {Scoglio},\ and\
  \citenamefont {Albin}}]{shakeri2016generalized}%
  \BibitemOpen
  \bibfield  {author} {\bibinfo {author} {\bibfnamefont {H.}~\bibnamefont
  {Shakeri}}, \bibinfo {author} {\bibfnamefont {P.}~\bibnamefont
  {Poggi-Corradini}}, \bibinfo {author} {\bibfnamefont {C.}~\bibnamefont
  {Scoglio}}, \ and\ \bibinfo {author} {\bibfnamefont {N.}~\bibnamefont
  {Albin}},\ }\href@noop {} {\bibfield  {journal} {\bibinfo  {journal} {Journal
  of Computational and Applied Mathematics}\ } (\bibinfo {year}
  {2016})}\BibitemShut {NoStop}%
\bibitem [{\citenamefont {Goering}\ \emph {et~al.}(2015)\citenamefont
  {Goering}, \citenamefont {Sahneh}, \citenamefont {Albin}, \citenamefont
  {Scoglio},\ and\ \citenamefont {Poggi-Corradini}}]{goering2015numerical}%
  \BibitemOpen
  \bibfield  {author} {\bibinfo {author} {\bibfnamefont {M.}~\bibnamefont
  {Goering}}, \bibinfo {author} {\bibfnamefont {F.~D.}\ \bibnamefont {Sahneh}},
  \bibinfo {author} {\bibfnamefont {N.}~\bibnamefont {Albin}}, \bibinfo
  {author} {\bibfnamefont {C.}~\bibnamefont {Scoglio}}, \ and\ \bibinfo
  {author} {\bibfnamefont {P.}~\bibnamefont {Poggi-Corradini}},\ }\href@noop {}
  {\bibfield  {journal} {\bibinfo  {journal} {arXiv preprint arXiv:1511.07893}\
  } (\bibinfo {year} {2015})}\BibitemShut {NoStop}%
\bibitem [{\citenamefont {Boyd}\ and\ \citenamefont
  {Vandenberghe}(2004)}]{boyd2004convex}%
  \BibitemOpen
  \bibfield  {author} {\bibinfo {author} {\bibfnamefont {S.}~\bibnamefont
  {Boyd}}\ and\ \bibinfo {author} {\bibfnamefont {L.}~\bibnamefont
  {Vandenberghe}},\ }\href@noop {} {\enquote {\bibinfo {title} {Convex
  optimization},}\ } (\bibinfo {year} {2004})\BibitemShut {NoStop}%
\bibitem [{\citenamefont {Goldfarb}\ and\ \citenamefont
  {Idnani}(1983)}]{goldfarb1983numerically}%
  \BibitemOpen
  \bibfield  {author} {\bibinfo {author} {\bibfnamefont {D.}~\bibnamefont
  {Goldfarb}}\ and\ \bibinfo {author} {\bibfnamefont {A.}~\bibnamefont
  {Idnani}},\ }\href@noop {} {\bibfield  {journal} {\bibinfo  {journal}
  {Mathematical programming}\ }\textbf {\bibinfo {volume} {27}},\ \bibinfo
  {pages} {1} (\bibinfo {year} {1983})}\BibitemShut {NoStop}%
\bibitem [{\citenamefont {Watts}\ and\ \citenamefont
  {Strogatz}(1998)}]{watts1998collective}%
  \BibitemOpen
  \bibfield  {author} {\bibinfo {author} {\bibfnamefont {D.~J.}\ \bibnamefont
  {Watts}}\ and\ \bibinfo {author} {\bibfnamefont {S.~H.}\ \bibnamefont
  {Strogatz}},\ }\href@noop {} {\bibfield  {journal} {\bibinfo  {journal}
  {nature}\ }\textbf {\bibinfo {volume} {393}},\ \bibinfo {pages} {440}
  (\bibinfo {year} {1998})}\BibitemShut {NoStop}%
\bibitem [{\citenamefont {Barabasi}\ and\ \citenamefont
  {Albert}(1999)}]{Barabasi99emergenceScaling}%
  \BibitemOpen
  \bibfield  {author} {\bibinfo {author} {\bibfnamefont {A.}~\bibnamefont
  {Barabasi}}\ and\ \bibinfo {author} {\bibfnamefont {R.}~\bibnamefont
  {Albert}},\ }\href@noop {} {\bibfield  {journal} {\bibinfo  {journal}
  {Science}\ }\textbf {\bibinfo {volume} {286}},\ \bibinfo {pages} {509}
  (\bibinfo {year} {1999})},\ \Eprint
  {http://arxiv.org/abs/http://www.sciencemag.org/cgi/reprint/286/5439/509.pdf}
  {http://www.sciencemag.org/cgi/reprint/286/5439/509.pdf} \BibitemShut
  {NoStop}%
\bibitem [{\citenamefont {Newman}(2003{\natexlab{b}})}]{newman2003ego}%
  \BibitemOpen
  \bibfield  {author} {\bibinfo {author} {\bibfnamefont {M.~E.}\ \bibnamefont
  {Newman}},\ }\href@noop {} {\bibfield  {journal} {\bibinfo  {journal} {Social
  Networks}\ }\textbf {\bibinfo {volume} {25}},\ \bibinfo {pages} {83}
  (\bibinfo {year} {2003}{\natexlab{b}})}\BibitemShut {NoStop}%
\bibitem [{\citenamefont {Caldarelli}\ \emph {et~al.}(2004)\citenamefont
  {Caldarelli}, \citenamefont {Pastor-Satorras},\ and\ \citenamefont
  {Vespignani}}]{caldarelli2004structure}%
  \BibitemOpen
  \bibfield  {author} {\bibinfo {author} {\bibfnamefont {G.}~\bibnamefont
  {Caldarelli}}, \bibinfo {author} {\bibfnamefont {R.}~\bibnamefont
  {Pastor-Satorras}}, \ and\ \bibinfo {author} {\bibfnamefont {A.}~\bibnamefont
  {Vespignani}},\ }\href@noop {} {\bibfield  {journal} {\bibinfo  {journal}
  {The European Physical Journal B-Condensed Matter and Complex Systems}\
  }\textbf {\bibinfo {volume} {38}},\ \bibinfo {pages} {183} (\bibinfo {year}
  {2004})}\BibitemShut {NoStop}%
\bibitem [{\citenamefont {Lind}\ and\ \citenamefont
  {Herrmann}(2007)}]{lind2007new}%
  \BibitemOpen
  \bibfield  {author} {\bibinfo {author} {\bibfnamefont {P.~G.}\ \bibnamefont
  {Lind}}\ and\ \bibinfo {author} {\bibfnamefont {H.~J.}\ \bibnamefont
  {Herrmann}},\ }\href@noop {} {\bibfield  {journal} {\bibinfo  {journal} {New
  Journal of Physics}\ }\textbf {\bibinfo {volume} {9}},\ \bibinfo {pages}
  {228} (\bibinfo {year} {2007})}\BibitemShut {NoStop}%
\bibitem [{\citenamefont {Fronczak}\ \emph {et~al.}(2002)\citenamefont
  {Fronczak}, \citenamefont {Ho{\l}yst}, \citenamefont {Jedynak},\ and\
  \citenamefont {Sienkiewicz}}]{fronczak2002higher}%
  \BibitemOpen
  \bibfield  {author} {\bibinfo {author} {\bibfnamefont {A.}~\bibnamefont
  {Fronczak}}, \bibinfo {author} {\bibfnamefont {J.~A.}\ \bibnamefont
  {Ho{\l}yst}}, \bibinfo {author} {\bibfnamefont {M.}~\bibnamefont {Jedynak}},
  \ and\ \bibinfo {author} {\bibfnamefont {J.}~\bibnamefont {Sienkiewicz}},\
  }\href@noop {} {\bibfield  {journal} {\bibinfo  {journal} {Physica A:
  Statistical Mechanics and its Applications}\ }\textbf {\bibinfo {volume}
  {316}},\ \bibinfo {pages} {688} (\bibinfo {year} {2002})}\BibitemShut
  {NoStop}%
\bibitem [{\citenamefont {Soffer}\ and\ \citenamefont
  {Vazquez}(2005)}]{soffer2005network}%
  \BibitemOpen
  \bibfield  {author} {\bibinfo {author} {\bibfnamefont {S.~N.}\ \bibnamefont
  {Soffer}}\ and\ \bibinfo {author} {\bibfnamefont {A.}~\bibnamefont
  {Vazquez}},\ }\href@noop {} {\bibfield  {journal} {\bibinfo  {journal}
  {Physical Review E}\ }\textbf {\bibinfo {volume} {71}},\ \bibinfo {pages}
  {057101} (\bibinfo {year} {2005})}\BibitemShut {NoStop}%
\bibitem [{\citenamefont {Saram{\"a}ki}\ \emph {et~al.}(2007)\citenamefont
  {Saram{\"a}ki}, \citenamefont {Kivel{\"a}}, \citenamefont {Onnela},
  \citenamefont {Kaski},\ and\ \citenamefont
  {Kertesz}}]{saramaki2007generalizations}%
  \BibitemOpen
  \bibfield  {author} {\bibinfo {author} {\bibfnamefont {J.}~\bibnamefont
  {Saram{\"a}ki}}, \bibinfo {author} {\bibfnamefont {M.}~\bibnamefont
  {Kivel{\"a}}}, \bibinfo {author} {\bibfnamefont {J.-P.}\ \bibnamefont
  {Onnela}}, \bibinfo {author} {\bibfnamefont {K.}~\bibnamefont {Kaski}}, \
  and\ \bibinfo {author} {\bibfnamefont {J.}~\bibnamefont {Kertesz}},\
  }\href@noop {} {\bibfield  {journal} {\bibinfo  {journal} {Physical Review
  E}\ }\textbf {\bibinfo {volume} {75}},\ \bibinfo {pages} {027105} (\bibinfo
  {year} {2007})}\BibitemShut {NoStop}%
\bibitem [{\citenamefont {Opsahl}\ and\ \citenamefont
  {Panzarasa}(2009)}]{opsahl2009clustering}%
  \BibitemOpen
  \bibfield  {author} {\bibinfo {author} {\bibfnamefont {T.}~\bibnamefont
  {Opsahl}}\ and\ \bibinfo {author} {\bibfnamefont {P.}~\bibnamefont
  {Panzarasa}},\ }\href@noop {} {\bibfield  {journal} {\bibinfo  {journal}
  {Social networks}\ }\textbf {\bibinfo {volume} {31}},\ \bibinfo {pages} {155}
  (\bibinfo {year} {2009})}\BibitemShut {NoStop}%
\bibitem [{\citenamefont {Gleiser}\ and\ \citenamefont {Danon}(2003)}]{Jazz}%
  \BibitemOpen
  \bibfield  {author} {\bibinfo {author} {\bibfnamefont {P.~M.}\ \bibnamefont
  {Gleiser}}\ and\ \bibinfo {author} {\bibfnamefont {L.}~\bibnamefont
  {Danon}},\ }\href@noop {} {\bibfield  {journal} {\bibinfo  {journal}
  {Advances in complex systems}\ }\textbf {\bibinfo {volume} {6}},\ \bibinfo
  {pages} {565} (\bibinfo {year} {2003})}\BibitemShut {NoStop}%
\bibitem [{\citenamefont {Guimera}\ \emph {et~al.}(2003)\citenamefont
  {Guimera}, \citenamefont {Danon}, \citenamefont {Diaz-Guilera}, \citenamefont
  {Giralt},\ and\ \citenamefont {Arenas}}]{guimera2003self}%
  \BibitemOpen
  \bibfield  {author} {\bibinfo {author} {\bibfnamefont {R.}~\bibnamefont
  {Guimera}}, \bibinfo {author} {\bibfnamefont {L.}~\bibnamefont {Danon}},
  \bibinfo {author} {\bibfnamefont {A.}~\bibnamefont {Diaz-Guilera}}, \bibinfo
  {author} {\bibfnamefont {F.}~\bibnamefont {Giralt}}, \ and\ \bibinfo {author}
  {\bibfnamefont {A.}~\bibnamefont {Arenas}},\ }\href@noop {} {\bibfield
  {journal} {\bibinfo  {journal} {Physical review E}\ }\textbf {\bibinfo
  {volume} {68}},\ \bibinfo {pages} {065103} (\bibinfo {year}
  {2003})}\BibitemShut {NoStop}%
\bibitem [{\citenamefont {Kunegis}(2014)}]{kunegis2014handbook}%
  \BibitemOpen
  \bibfield  {author} {\bibinfo {author} {\bibfnamefont {J.}~\bibnamefont
  {Kunegis}},\ }\href@noop {} {\bibfield  {journal} {\bibinfo  {journal} {arXiv
  preprint arXiv:1402.5500}\ } (\bibinfo {year} {2014})}\BibitemShut {NoStop}%
\bibitem [{\citenamefont {McAuley}\ and\ \citenamefont
  {Leskovec}(2012)}]{mcauley2012learning}%
  \BibitemOpen
  \bibfield  {author} {\bibinfo {author} {\bibfnamefont {J.~J.}\ \bibnamefont
  {McAuley}}\ and\ \bibinfo {author} {\bibfnamefont {J.}~\bibnamefont
  {Leskovec}},\ }in\ \href@noop {} {\emph {\bibinfo {booktitle} {NIPS}}},\
  Vol.\ \bibinfo {volume} {2012}\ (\bibinfo {year} {2012})\ pp.\ \bibinfo
  {pages} {548--56}\BibitemShut {NoStop}%
\bibitem [{Ham()}]{Hamster}%
  \BibitemOpen
  \href@noop {} {\enquote {\bibinfo {title} {{ Hamsterster friendships network
  dataset, KONECT}},}\ }\bibinfo {howpublished}
  {\url{http://konect.uni-koblenz.de/networks/petster-friendships-hamster}},\
  \bibinfo {note} {accessed: 2016-08-11}\BibitemShut {NoStop}%
\bibitem [{\citenamefont {Homans}(2013)}]{homans2013human}%
  \BibitemOpen
  \bibfield  {author} {\bibinfo {author} {\bibfnamefont {G.~C.}\ \bibnamefont
  {Homans}},\ }\href@noop {} {\emph {\bibinfo {title} {The human group}}},\
  Vol.~\bibinfo {volume} {7}\ (\bibinfo  {publisher} {Routledge},\ \bibinfo
  {year} {2013})\BibitemShut {NoStop}%
\bibitem [{\citenamefont {Ravasz}\ \emph {et~al.}(2002)\citenamefont {Ravasz},
  \citenamefont {Somera}, \citenamefont {Mongru}, \citenamefont {Oltvai},\ and\
  \citenamefont {Barab{\'a}si}}]{ravasz2002hierarchical}%
  \BibitemOpen
  \bibfield  {author} {\bibinfo {author} {\bibfnamefont {E.}~\bibnamefont
  {Ravasz}}, \bibinfo {author} {\bibfnamefont {A.~L.}\ \bibnamefont {Somera}},
  \bibinfo {author} {\bibfnamefont {D.~A.}\ \bibnamefont {Mongru}}, \bibinfo
  {author} {\bibfnamefont {Z.~N.}\ \bibnamefont {Oltvai}}, \ and\ \bibinfo
  {author} {\bibfnamefont {A.-L.}\ \bibnamefont {Barab{\'a}si}},\ }\href@noop
  {} {\bibfield  {journal} {\bibinfo  {journal} {science}\ }\textbf {\bibinfo
  {volume} {297}},\ \bibinfo {pages} {1551} (\bibinfo {year}
  {2002})}\BibitemShut {NoStop}%
\bibitem [{\citenamefont {Fortunato}(2010)}]{fortunato2010community}%
  \BibitemOpen
  \bibfield  {author} {\bibinfo {author} {\bibfnamefont {S.}~\bibnamefont
  {Fortunato}},\ }\href@noop {} {\bibfield  {journal} {\bibinfo  {journal}
  {Physics reports}\ }\textbf {\bibinfo {volume} {486}},\ \bibinfo {pages} {75}
  (\bibinfo {year} {2010})}\BibitemShut {NoStop}%
\bibitem [{\citenamefont {Radicchi}\ \emph {et~al.}(2004)\citenamefont
  {Radicchi}, \citenamefont {Castellano}, \citenamefont {Cecconi},
  \citenamefont {Loreto},\ and\ \citenamefont {Parisi}}]{radicchi2004defining}%
  \BibitemOpen
  \bibfield  {author} {\bibinfo {author} {\bibfnamefont {F.}~\bibnamefont
  {Radicchi}}, \bibinfo {author} {\bibfnamefont {C.}~\bibnamefont
  {Castellano}}, \bibinfo {author} {\bibfnamefont {F.}~\bibnamefont {Cecconi}},
  \bibinfo {author} {\bibfnamefont {V.}~\bibnamefont {Loreto}}, \ and\ \bibinfo
  {author} {\bibfnamefont {D.}~\bibnamefont {Parisi}},\ }\href@noop {}
  {\bibfield  {journal} {\bibinfo  {journal} {Proceedings of the National
  Academy of Sciences of the United States of America}\ }\textbf {\bibinfo
  {volume} {101}},\ \bibinfo {pages} {2658} (\bibinfo {year}
  {2004})}\BibitemShut {NoStop}%
\bibitem [{\citenamefont {Vragovi{\'c}}\ and\ \citenamefont
  {Louis}(2006)}]{vragovic2006network}%
  \BibitemOpen
  \bibfield  {author} {\bibinfo {author} {\bibfnamefont {I.}~\bibnamefont
  {Vragovi{\'c}}}\ and\ \bibinfo {author} {\bibfnamefont {E.}~\bibnamefont
  {Louis}},\ }\href@noop {} {\bibfield  {journal} {\bibinfo  {journal}
  {Physical Review E}\ }\textbf {\bibinfo {volume} {74}},\ \bibinfo {pages}
  {016105} (\bibinfo {year} {2006})}\BibitemShut {NoStop}%
\bibitem [{\citenamefont {Berry}\ \emph {et~al.}(2011)\citenamefont {Berry},
  \citenamefont {Hendrickson}, \citenamefont {LaViolette},\ and\ \citenamefont
  {Phillips}}]{berry2011tolerating}%
  \BibitemOpen
  \bibfield  {author} {\bibinfo {author} {\bibfnamefont {J.~W.}\ \bibnamefont
  {Berry}}, \bibinfo {author} {\bibfnamefont {B.}~\bibnamefont {Hendrickson}},
  \bibinfo {author} {\bibfnamefont {R.~A.}\ \bibnamefont {LaViolette}}, \ and\
  \bibinfo {author} {\bibfnamefont {C.~A.}\ \bibnamefont {Phillips}},\
  }\href@noop {} {\bibfield  {journal} {\bibinfo  {journal} {Physical Review
  E}\ }\textbf {\bibinfo {volume} {83}},\ \bibinfo {pages} {056119} (\bibinfo
  {year} {2011})}\BibitemShut {NoStop}%
\bibitem [{\citenamefont {Fortunato}\ and\ \citenamefont
  {Barthelemy}(2007)}]{fortunato2007resolution}%
  \BibitemOpen
  \bibfield  {author} {\bibinfo {author} {\bibfnamefont {S.}~\bibnamefont
  {Fortunato}}\ and\ \bibinfo {author} {\bibfnamefont {M.}~\bibnamefont
  {Barthelemy}},\ }\href@noop {} {\bibfield  {journal} {\bibinfo  {journal}
  {Proceedings of the National Academy of Sciences}\ }\textbf {\bibinfo
  {volume} {104}},\ \bibinfo {pages} {36} (\bibinfo {year} {2007})}\BibitemShut
  {NoStop}%
\bibitem [{\citenamefont {Khadivi}\ \emph {et~al.}(2011)\citenamefont
  {Khadivi}, \citenamefont {Rad},\ and\ \citenamefont
  {Hasler}}]{khadivi2011network}%
  \BibitemOpen
  \bibfield  {author} {\bibinfo {author} {\bibfnamefont {A.}~\bibnamefont
  {Khadivi}}, \bibinfo {author} {\bibfnamefont {A.~A.}\ \bibnamefont {Rad}}, \
  and\ \bibinfo {author} {\bibfnamefont {M.}~\bibnamefont {Hasler}},\
  }\href@noop {} {\bibfield  {journal} {\bibinfo  {journal} {Physical Review
  E}\ }\textbf {\bibinfo {volume} {83}},\ \bibinfo {pages} {046104} (\bibinfo
  {year} {2011})}\BibitemShut {NoStop}%
\bibitem [{\citenamefont {Freeman}(1977)}]{freeman1977set}%
  \BibitemOpen
  \bibfield  {author} {\bibinfo {author} {\bibfnamefont {L.~C.}\ \bibnamefont
  {Freeman}},\ }\href@noop {} {\bibfield  {journal} {\bibinfo  {journal}
  {Sociometry}\ ,\ \bibinfo {pages} {35}} (\bibinfo {year} {1977})}\BibitemShut
  {NoStop}%
\bibitem [{\citenamefont {Malliaros}\ and\ \citenamefont
  {Vazirgiannis}(2013)}]{malliaros2013clustering}%
  \BibitemOpen
  \bibfield  {author} {\bibinfo {author} {\bibfnamefont {F.~D.}\ \bibnamefont
  {Malliaros}}\ and\ \bibinfo {author} {\bibfnamefont {M.}~\bibnamefont
  {Vazirgiannis}},\ }\href@noop {} {\bibfield  {journal} {\bibinfo  {journal}
  {Physics Reports}\ }\textbf {\bibinfo {volume} {533}},\ \bibinfo {pages} {95}
  (\bibinfo {year} {2013})}\BibitemShut {NoStop}%
\bibitem [{\citenamefont {Klymko}\ \emph {et~al.}(2014)\citenamefont {Klymko},
  \citenamefont {Gleich},\ and\ \citenamefont {Kolda}}]{klymko2014using}%
  \BibitemOpen
  \bibfield  {author} {\bibinfo {author} {\bibfnamefont {C.}~\bibnamefont
  {Klymko}}, \bibinfo {author} {\bibfnamefont {D.}~\bibnamefont {Gleich}}, \
  and\ \bibinfo {author} {\bibfnamefont {T.~G.}\ \bibnamefont {Kolda}},\
  }\href@noop {} {\bibfield  {journal} {\bibinfo  {journal} {arXiv preprint
  arXiv:1404.5874}\ } (\bibinfo {year} {2014})}\BibitemShut {NoStop}%
\bibitem [{\citenamefont {Zachary}(1977)}]{ZacharyKarate}%
  \BibitemOpen
  \bibfield  {author} {\bibinfo {author} {\bibfnamefont {W.~W.}\ \bibnamefont
  {Zachary}},\ }\href@noop {} {\bibfield  {journal} {\bibinfo  {journal}
  {Journal of anthropological research}\ ,\ \bibinfo {pages} {452}} (\bibinfo
  {year} {1977})}\BibitemShut {NoStop}%
\bibitem [{\citenamefont {Barab{\'a}si}(2013)}]{barabasi2013network}%
  \BibitemOpen
  \bibfield  {author} {\bibinfo {author} {\bibfnamefont {A.-L.}\ \bibnamefont
  {Barab{\'a}si}},\ }\href@noop {} {\bibfield  {journal} {\bibinfo  {journal}
  {Philosophical Transactions of the Royal Society of London A: Mathematical,
  Physical and Engineering Sciences}\ }\textbf {\bibinfo {volume} {371}},\
  \bibinfo {pages} {20120375} (\bibinfo {year} {2013})}\BibitemShut {NoStop}%
\bibitem [{\citenamefont {Newman}(2010)}]{NewmanBook}%
  \BibitemOpen
  \bibfield  {author} {\bibinfo {author} {\bibfnamefont {M.~E.~J.}\
  \bibnamefont {Newman}},\ }\href@noop {} {\emph {\bibinfo {title} {Networks:
  An Introduction}}}\ (\bibinfo  {publisher} {Oxford},\ \bibinfo {year}
  {2010})\BibitemShut {NoStop}%
\bibitem [{\citenamefont {Palla}\ \emph {et~al.}(2007)\citenamefont {Palla},
  \citenamefont {Barab{\'a}si},\ and\ \citenamefont
  {Vicsek}}]{palla2007quantifying}%
  \BibitemOpen
  \bibfield  {author} {\bibinfo {author} {\bibfnamefont {G.}~\bibnamefont
  {Palla}}, \bibinfo {author} {\bibfnamefont {A.-L.}\ \bibnamefont
  {Barab{\'a}si}}, \ and\ \bibinfo {author} {\bibfnamefont {T.}~\bibnamefont
  {Vicsek}},\ }\href@noop {} {\bibfield  {journal} {\bibinfo  {journal}
  {Nature}\ }\textbf {\bibinfo {volume} {446}},\ \bibinfo {pages} {664}
  (\bibinfo {year} {2007})}\BibitemShut {NoStop}%
\bibitem [{\citenamefont {Palla}\ \emph {et~al.}(2005)\citenamefont {Palla},
  \citenamefont {Der{\'e}nyi}, \citenamefont {Farkas},\ and\ \citenamefont
  {Vicsek}}]{Palla}%
  \BibitemOpen
  \bibfield  {author} {\bibinfo {author} {\bibfnamefont {G.}~\bibnamefont
  {Palla}}, \bibinfo {author} {\bibfnamefont {I.}~\bibnamefont {Der{\'e}nyi}},
  \bibinfo {author} {\bibfnamefont {I.}~\bibnamefont {Farkas}}, \ and\ \bibinfo
  {author} {\bibfnamefont {T.}~\bibnamefont {Vicsek}},\ }\href@noop {}
  {\bibfield  {journal} {\bibinfo  {journal} {Nature}\ }\textbf {\bibinfo
  {volume} {435}},\ \bibinfo {pages} {814} (\bibinfo {year}
  {2005})}\BibitemShut {NoStop}%
\bibitem [{\citenamefont {Clauset}\ \emph {et~al.}(2004)\citenamefont
  {Clauset}, \citenamefont {Newman},\ and\ \citenamefont
  {Moore}}]{clauset2004finding}%
  \BibitemOpen
  \bibfield  {author} {\bibinfo {author} {\bibfnamefont {A.}~\bibnamefont
  {Clauset}}, \bibinfo {author} {\bibfnamefont {M.~E.}\ \bibnamefont {Newman}},
  \ and\ \bibinfo {author} {\bibfnamefont {C.}~\bibnamefont {Moore}},\
  }\href@noop {} {\bibfield  {journal} {\bibinfo  {journal} {Physical review
  E}\ }\textbf {\bibinfo {volume} {70}},\ \bibinfo {pages} {066111} (\bibinfo
  {year} {2004})}\BibitemShut {NoStop}%
\bibitem [{\citenamefont {Blondel}\ \emph {et~al.}(2008)\citenamefont
  {Blondel}, \citenamefont {Guillaume}, \citenamefont {Lambiotte},\ and\
  \citenamefont {Lefebvre}}]{blondel2008fast}%
  \BibitemOpen
  \bibfield  {author} {\bibinfo {author} {\bibfnamefont {V.~D.}\ \bibnamefont
  {Blondel}}, \bibinfo {author} {\bibfnamefont {J.-L.}\ \bibnamefont
  {Guillaume}}, \bibinfo {author} {\bibfnamefont {R.}~\bibnamefont
  {Lambiotte}}, \ and\ \bibinfo {author} {\bibfnamefont {E.}~\bibnamefont
  {Lefebvre}},\ }\href@noop {} {\bibfield  {journal} {\bibinfo  {journal}
  {Journal of statistical mechanics: theory and experiment}\ }\textbf {\bibinfo
  {volume} {2008}},\ \bibinfo {pages} {P10008} (\bibinfo {year}
  {2008})}\BibitemShut {NoStop}%
\bibitem [{\citenamefont {Lancichinetti}\ \emph {et~al.}(2008)\citenamefont
  {Lancichinetti}, \citenamefont {Fortunato},\ and\ \citenamefont
  {Radicchi}}]{lancichinetti2008benchmark}%
  \BibitemOpen
  \bibfield  {author} {\bibinfo {author} {\bibfnamefont {A.}~\bibnamefont
  {Lancichinetti}}, \bibinfo {author} {\bibfnamefont {S.}~\bibnamefont
  {Fortunato}}, \ and\ \bibinfo {author} {\bibfnamefont {F.}~\bibnamefont
  {Radicchi}},\ }\href@noop {} {\bibfield  {journal} {\bibinfo  {journal}
  {Physical review E}\ }\textbf {\bibinfo {volume} {78}},\ \bibinfo {pages}
  {046110} (\bibinfo {year} {2008})}\BibitemShut {NoStop}%
\bibitem [{\citenamefont {Girvan}\ and\ \citenamefont
  {Newman}(2002)}]{girvan2002community}%
  \BibitemOpen
  \bibfield  {author} {\bibinfo {author} {\bibfnamefont {M.}~\bibnamefont
  {Girvan}}\ and\ \bibinfo {author} {\bibfnamefont {M.~E.}\ \bibnamefont
  {Newman}},\ }\href@noop {} {\bibfield  {journal} {\bibinfo  {journal}
  {Proceedings of the national academy of sciences}\ }\textbf {\bibinfo
  {volume} {99}},\ \bibinfo {pages} {7821} (\bibinfo {year}
  {2002})}\BibitemShut {NoStop}%
\bibitem [{\citenamefont {Danon}\ \emph {et~al.}(2005)\citenamefont {Danon},
  \citenamefont {Diaz-Guilera}, \citenamefont {Duch},\ and\ \citenamefont
  {Arenas}}]{danon2005comparing}%
  \BibitemOpen
  \bibfield  {author} {\bibinfo {author} {\bibfnamefont {L.}~\bibnamefont
  {Danon}}, \bibinfo {author} {\bibfnamefont {A.}~\bibnamefont {Diaz-Guilera}},
  \bibinfo {author} {\bibfnamefont {J.}~\bibnamefont {Duch}}, \ and\ \bibinfo
  {author} {\bibfnamefont {A.}~\bibnamefont {Arenas}},\ }\href@noop {}
  {\bibfield  {journal} {\bibinfo  {journal} {Journal of Statistical Mechanics:
  Theory and Experiment}\ }\textbf {\bibinfo {volume} {2005}},\ \bibinfo
  {pages} {P09008} (\bibinfo {year} {2005})}\BibitemShut {NoStop}%
\bibitem [{\citenamefont {Li}\ \emph {et~al.}(2010)\citenamefont {Li},
  \citenamefont {Duan}, \citenamefont {Chen},\ and\ \citenamefont
  {Huang}}]{li2010consensus}%
  \BibitemOpen
  \bibfield  {author} {\bibinfo {author} {\bibfnamefont {Z.}~\bibnamefont
  {Li}}, \bibinfo {author} {\bibfnamefont {Z.}~\bibnamefont {Duan}}, \bibinfo
  {author} {\bibfnamefont {G.}~\bibnamefont {Chen}}, \ and\ \bibinfo {author}
  {\bibfnamefont {L.}~\bibnamefont {Huang}},\ }\href@noop {} {\bibfield
  {journal} {\bibinfo  {journal} {IEEE Transactions on Circuits and Systems I:
  Regular Papers}\ }\textbf {\bibinfo {volume} {57}},\ \bibinfo {pages} {213}
  (\bibinfo {year} {2010})}\BibitemShut {NoStop}%
\bibitem [{\citenamefont {Kuramoto}(1975)}]{kuramoto1975self}%
  \BibitemOpen
  \bibfield  {author} {\bibinfo {author} {\bibfnamefont {Y.}~\bibnamefont
  {Kuramoto}},\ }in\ \href@noop {} {\emph {\bibinfo {booktitle} {International
  symposium on mathematical problems in theoretical physics}}}\ (\bibinfo
  {organization} {Springer},\ \bibinfo {year} {1975})\ pp.\ \bibinfo {pages}
  {420--422}\BibitemShut {NoStop}%
\bibitem [{\citenamefont {Van~Mieghem}(2011)}]{van2011n}%
  \BibitemOpen
  \bibfield  {author} {\bibinfo {author} {\bibfnamefont {P.}~\bibnamefont
  {Van~Mieghem}},\ }\href@noop {} {\bibfield  {journal} {\bibinfo  {journal}
  {Computing}\ }\textbf {\bibinfo {volume} {93}},\ \bibinfo {pages} {147}
  (\bibinfo {year} {2011})}\BibitemShut {NoStop}%
\bibitem [{\citenamefont {Butts}(2000)}]{butts2000axiomatic}%
  \BibitemOpen
  \bibfield  {author} {\bibinfo {author} {\bibfnamefont {C.~T.}\ \bibnamefont
  {Butts}},\ }\href@noop {} {\bibfield  {journal} {\bibinfo  {journal} {Journal
  of Mathematical Sociology}\ }\textbf {\bibinfo {volume} {24}},\ \bibinfo
  {pages} {273} (\bibinfo {year} {2000})}\BibitemShut {NoStop}%
\end{thebibliography}%

\end{document}